\newif\ifpreprint 
\preprinttrue   
\ifpreprint
    \documentclass[manuscript,screen,nonacm]{acmart}
\else
    \documentclass[manuscript,review,anonymous]{acmart}
\fi

\AtBeginDocument{%
  }

\ifpreprint
    \setcopyright{none}
\else
    \setcopyright{acmlicensed}
    \copyrightyear{2027}
    \acmYear{2027}
    \acmDOI{XXXXXXX.XXXXXXX}
    \acmConference[Conference acronym '27]{Make sure to enter the correct
      conference title from your rights confirmation email}{June 03--05,
      2018}{Woodstock, NY}
    \acmISBN{978-1-4503-XXXX-X/2018/06}
\fi

\usepackage{xcolor}          
\usepackage{graphicx}
\usepackage{textcomp}         
\usepackage{listings}
\usepackage{xspace}
\usepackage{multirow}
\usepackage{colortbl}
\usepackage{wrapfig}
\usepackage{subcaption}
\usepackage{balance}
\usepackage[acronym]{glossaries}
\usepackage{booktabs}
\usepackage{array}
\ifpreprint
  \usepackage[textsize=scriptsize,disable]{todonotes}
\else
  \usepackage[textsize=scriptsize]{todonotes}
\fi
\usepackage{hyperref}         
 
\usepackage[T1]{fontenc}
\usepackage[normalem]{ulem}   
\usepackage{soul}             

\usepackage{tikz}
\usetikzlibrary{arrows.meta, positioning, shapes.geometric, calc, shadows}

\makeglossaries

\definecolor{hhcolor}{HTML}{2166AC}      
\definecolor{haicolor}{HTML}{C65500}     
\definecolor{aiaicolor}{HTML}{B2182B}    

\definecolor{rolestatecolor}{HTML}{D55E00}    
\definecolor{contextcolor}{HTML}{0072B2}      
\definecolor{orientationcolor}{HTML}{CC79A7}  
\definecolor{supporttypecolor}{HTML}{009E73}  
\definecolor{rolealignmentcolor}{HTML}{D55E00}

\colorlet{tableheadcolor}{gray!25}
\colorlet{tablerowcolor}{gray!15}
\colorlet{tablerowcolor2}{gray!30}

\definecolor{codegreen}{rgb}{0,0.6,0}
\definecolor{codegray}{rgb}{0.5,0.5,0.5}
\definecolor{codepurple}{rgb}{0.58,0,0.82}
\definecolor{backcolour}{rgb}{0.95,0.95,0.92}

\definecolor{lightblue}{rgb}{0.53, 0.81, 0.98}
\definecolor{darkblue}{rgb}{0.0, 0.0, 0.55}
\definecolor{elecblue}{rgb}{0.0, 0.5, 1.0}
\definecolor{lightgreen}{rgb}{0.56, 0.93, 0.56}
\definecolor{lightred}{rgb}{1.0, 0.71, 0.76}
\definecolor{lightorange}{rgb}{1.0, 0.87, 0.68}
\definecolor{lightpurple}{rgb}{0.87, 0.63, 0.87}
\definecolor{editcol}{HTML}{1E90FF}

\newcommand{\hh}{\textcolor{hhcolor}{\textsf{H-H}}\xspace}
\newcommand{\hai}{\textcolor{haicolor}{\textsf{H-AI}}\xspace}
\newcommand{\aiai}{\textcolor{aiaicolor}{\textsf{AI-AI}}\xspace}

\newcommand{\rolestate}{\textcolor{rolestatecolor}{Role State}\xspace}
\newcommand{\rolestates}{\textcolor{rolestatecolor}{Role States}\xspace}

\newcommand{\code}[1]{%
  {\setlength{\fboxsep}{1pt}%
   \colorbox{backcolour}{\footnotesize\texttt{#1}}}}

\lstdefinestyle{mystyle}{%
  backgroundcolor=\color{backcolour},
  commentstyle=\color{codegreen},
  keywordstyle=\color{magenta},
  numberstyle=\tiny\color{codegray},
  stringstyle=\color{codepurple},
  basicstyle=\ttfamily\footnotesize,
  breakatwhitespace=false,
  breaklines=true,
  captionpos=b,
  keepspaces=true,
  numbers=left,
  numbersep=5pt,
  showspaces=false,
  showstringspaces=false,
  showtabs=false,
  tabsize=2
}
\newcommand{\para}[1]{\vspace{0.5em}\noindent\textbf{#1}~}

\newcommand{\note}[3]{\textcolor{#1}{[#3: #2]}}

\newcounter{mynote}[section]

\newcommand{\thenote}{\thesection.\arabic{mynote}}
\newcommand{\rnote}[1]{\textcolor{purple}{$\ll${RB~\thenote:} {\sf #1}$\gg$}}
 
\newcommand{\noteIT}[1]{\textcolor{lightorange}{$\ll${IT~\thenote:} {\sf #1}$\gg$}}
\ifpreprint
  \renewcommand{\note}[3]{}
  \renewcommand{\rnote}[1]{}
  \renewcommand{\noteIT}[1]{}
\fi

\newacronym{GT}{GaTech}{Georgia Institute of Technology}
\newacronym{MI}{MI}{Motivational Interviewing}
\newacronym{D2}{D2}{Orientation}
\newacronym{D3}{D3}{Support Type}

\newcommand{\AROMA}{\texttt{AROMA }\xspace}

\newcommand{\RoleExploring}{\textcolor{rolestatecolor}{\code{Exploring}}\xspace}
\newcommand{\RoleSuggesting}{\textcolor{rolestatecolor}{\code{Suggesting}}\xspace}
\newcommand{\RoleReflecting}{\textcolor{rolestatecolor}{\code{Reflecting}}\xspace}
\newcommand{\RoleFormulating}{\textcolor{rolestatecolor}{\code{Formulating}}\xspace}
\newcommand{\RoleAffirming}{\textcolor{rolestatecolor}{\code{Affirming}}\xspace}
\newcommand{\RoleRelating}{\textcolor{rolestatecolor}{\code{Relating}}\xspace}
\newcommand{\RoleReframing}{\textcolor{rolestatecolor}{\code{Reframing}}\xspace}
\newcommand{\RoleDirecting}{\textcolor{rolestatecolor}{\code{Directing}}\xspace}

\newcommand{\AskNur}{\texttt{Ask\texttimes Nur}\xspace}
\newcommand{\AskAF}{\texttt{Ask\texttimes AF}\xspace}
\newcommand{\RBNur}{\texttt{RB\texttimes Nur}\xspace}
\newcommand{\RBAF}{\texttt{RB\texttimes AF}\xspace}
\newcommand{\AckNur}{\texttt{Ack\texttimes Nur}\xspace}
\newcommand{\AckAF}{\texttt{Ack\texttimes AF}\xspace}
\newcommand{\AdvNur}{\texttt{Adv\texttimes Nur}\xspace}
\newcommand{\AdvAF}{\texttt{Adv\texttimes AF}\xspace}

\newcommand{\Asking}{\textit{Asking}\xspace}
\newcommand{\RB}{\textit{Repeating Back}\xspace}
\newcommand{\Ack}{\textit{Acknowledgement}\xspace}
\newcommand{\Adv}{\textit{Advising}\xspace}
\newcommand{\Nur}{\textit{Nurturant}\xspace}
\newcommand{\AF}{\textit{Action-Facilitating}\xspace}      
\author{Issac Ting}
\email{it2405@nyu.edu}
\orcid{0009-0009-4946-4731}
\affiliation{%
  \institution{NYU Tandon School of Engineering}
  \city{New York}
  \state{NY}
  \country{USA}
}

\author{Ethan Pan}
\email{pan248@wisc.edu}
\orcid{0009-0000-4676-4418}
\affiliation{%
  \institution{University of Wisconsin--Madison}
  \city{Madison}
  \state{WI}
  \country{USA}
}

\author{Koustuv Saha}
\email{ksaha2@illinois.edu}
\orcid{0000-0002-8872-2934}
\affiliation{%
  \institution{University of Illinois Urbana-Champaign}
  \city{Urbana}
  \state{IL}
  \country{USA}
}

\author{Hua Shen}
\email{hs3645@nyu.edu}
\orcid{0000-0002-4928-525X}
\affiliation{%
  \institution{NYU Shanghai}
  \city{Shanghai}
  \country{China}
}

\author{Kevin Chow}
\email{kkc7977@nyu.edu}
\orcid{0009-0008-5455-3328}
\affiliation{%
  \institution{NYU Tandon School of Engineering}
  \city{New York}
  \state{NY}
  \country{USA}
}

\author{Vedant Das Swain}
\email{vd2538@nyu.edu}
\orcid{0000-0001-6871-3523}
\affiliation{%
  \institution{NYU Tandon School of Engineering}
  \city{New York}
  \state{NY}
  \country{USA}
}       

\makeglossaries

\begin{document}

\title{AROMA: Understanding  AI Support For Mental Well-being through Role Theory}

\renewcommand{\shortauthors}{Ting et al.}

\begin{abstract}
AI supporters for mental well-being are becoming more prevalent. When these AI supporters start acting in ways beyond their expected function it leads to relational harm. Standard approaches to evaluating such AI supporters are limited to inspecting their efficacy in narrow task-specific settings, which do not explain these discrepancies in expected and actual behavior. To address this gap, we propose the AROMA (A Role-Oriented Model of Analysis) framework to provide a common role taxonomy for different support contexts. We developed AROMA using a theory-informed annotation of over 7,515 conversations from 13 datasets and propose eight "Role States" based on observed---not prescribed---behavior. We show these Role States vary according to the context of support; depict unique structural and temporal characteristics of AI supporters vis-a-vis Humans; and inform novel AI support prompting strategy for better role alignment. This framework provides a new lens to understand AI support failures.
\end{abstract}


 
\begin{teaserfigure}
    \centering
    \includegraphics[width=1\linewidth]{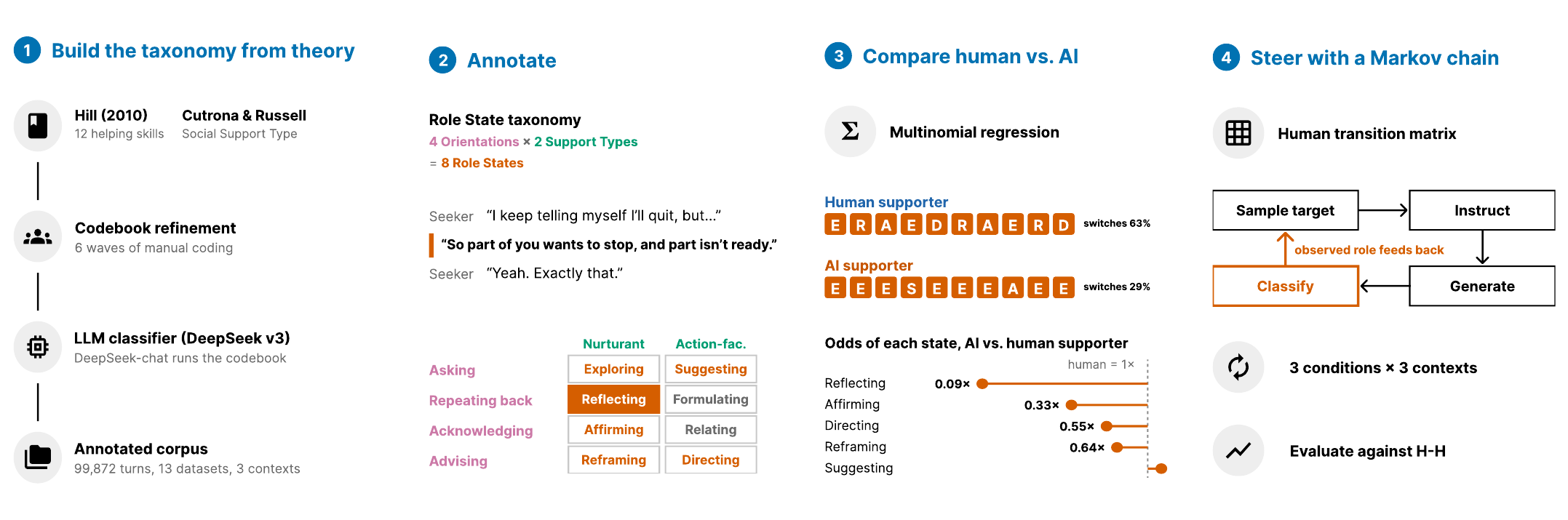}
    \caption{ (a) Using literature to ground taxonomy , iteratively coded using manual annotators before using LLM-as-a-judge to scale. (b) annotate using 2 set of labels to describe Role States. (c) using multinomial regression to validate statistical significance and odds ratio d) experimented with different ways to steer the support conversation using role probabilities.}
   \label{fig:teaser}
\end{teaserfigure} 

\maketitle


\section{Introduction}

LLM-based AI systems now routinely provide emotional support, counselling, and companionship. Millions of users turn to chatbots for mental health support, peer coaching, and romantic relationships~\cite{manoli_shes_2025, song_typing_2025}. This expansion from task assistance to relational care has redefined what users expect from conversational AI. 
Now, consider \textit{Tessa}, a wellness chatbot deployed by the National Eating Disorders Association in 2023~\cite{nytimes_tessa_2023}. Tessa was designed to play the role of a supportive companion that helps individuals cope with eating disorders. In practice, however, Tessa was found to be directing users into caloric restriction and weight monitoring---performing the role of a weight-loss coach. Several users found this behavior harmful and eventually Tessa was taken offline~\cite{nytimes_tessa_2023}. We know in traditional human support settings, to reduce harms it is critical that the role a seeker expects is aligned to the role a supporter performs~\cite{greenberg_are_2006}. Moreover, regardless of context, humans express roles in flexible and interchangeable ways \cite{biddle_recent_1986}. As we witness in Tessa's case, this collapse in roles does not only result in conversational error but in relational harm.


Although prior works have suggested approaches to evaluate AI chatbots in support settings, these methods have some key barriers that prevent it from understanding misalignment in roles. 
The dominant method of engineering the function of an AI supporter is through \textit{persona injection}, which is the practice of specifying a system's intended role via an initial system prompt \cite{xiao_when_2026}.
Several evaluation methods benchmark how AI perform their prescribed roles to measure if it can effectively meet its goals \cite{xiao_when_2026, samuel_personagym_2025}. These are useful to study the fidelity of support within a role, but does not account for scenarios that require a multitude of roles. Another set of methods focus on how AI behavior drifts from intended role~\cite{araujo_persistent_2026}, but does not concern itself with describing what roles the AI embodies instead. Thus, the current frameworks to assess AI support provides only narrow goal-specific understanding that is limited to specific domains. 

We posit an alternative approach to assessing AI support by drawing from literature on \textit{Role Theory}~\cite{biddle_recent_1986} that distinguishes \textit{prescribed role}---what an actor is assigned to do---from the \textit{enacted role}---what an actor actually does. This theory has been salient to understand interpersonal dynamics across a variety of domains ~\cite{biddle_recent_1986}. Despite its popularity, so far the literature does not describe roles through canonical labels (e.g., mentor) and thus does not provide clear lens to understand support behaviors in a transferrable way. Defining roles has always been challenging. Recently \citeauthor{lindgren_emerging_2025} stated that roles have shared properties across situations and can be ``partly overlapping and fluent'' depending on these situations and called for AI techniques that allows for "explainability, negotiation and adaptation of the enacted roles."~\cite{lindgren_emerging_2025} In response to this call, this paper presents \AROMA (A Role-Oriented Model of Analysis), a context-agnostic framework for understanding enacted roles in support conversations. 
We developed \AROMA by supplementing the theory in \textit{Role Theory} with empirical evidence from a corpora of human-human, human-AI, and AI-AI support conversations sourced from three different supper contexts peer support \cite{liuEmotionalSupportDialog2021, rashkin_towards_2019, yan_sharechat_2026, chen_socialsim_2025} , counselling \cite{malhotra_speaker_2022, qi_kokorochat_2025, qiu_interactive_2024, qiu_psydial_2025} and motivational interviewing~\cite{wu_anno-mi_2022, meyer_you_2024}. 
Using the \AROMA framework we answer the following questions: \textbf{[RQ1]} What features distinguish roles support-seeking contexts?; \textbf{[RQ2]} How does AI support roles differ from that of human actors?; and \textbf{[RQ3]} How can AI supporters be designed for better role alignment?   


We report how we developed \AROMA's taxonomy of \rolestates by annotating $7,515$ conversations from $13$ datasets using different dimensions of support---\textit{orientation} and \textit{support type}---that we derive from Hill's helping-skills model~\cite{hill_helping_2010} and Cutrona and Russell's optimal-matching theory~\cite{cutrona_type_1990}. We validate the taxonomy with an LLM-based classifier that achieved $\kappa = 0.734$ for orientation and $\mathrm{AC1} = 0.889$ for support type (\autoref{sec:methods}). Then our paper shows that support roles can be described through a shared taxonomy (\autoref{sec:rq1}). \RoleAffirming in peer support is the most prominent context-distinguishing state (\(OR = 5.35\)), alongside elevated \RoleReflecting in MI (\(OR = 2.25\)) and \RoleDirecting in peer support (\(OR = 3.18\)).  This shows how different context exhibited quantifiable different ways of support (RQ1). Next we present which can help assess the divergence in human and AI roles (\autoref{sec:rq2}) by observing changing in how frequently the role states change (\hh conversations switch roles on 62\% of consecutive turn pairs, compared with 36\% for \hai and 52\% for \aiai) . And finally, this informs how we might align them for different contexts . Steering reduces the distributional gap between AI and human role profiles by 70--85\% (\autoref{sec:rq3}).
Together our paper contributes the following:

\begin{itemize}
  \item A descriptive taxonomy of \rolestates operationalised by how supporters respond to seekers (\textit{orientation}) and what kind of support they provide (\textit{type}). These \rolestates are uniquely distributed across contexts, reinforcing how varying situations require different ways of support (RQ1).
  \item Empirical evidence of how AI support roles differs from humans in terms of AI's affinity towards certain roles as well as its probability of transitioning to different roles (RQ2).
  \item A Markov-informed steering mechanism that relies on role transition probabilities to provide per-turn scaffolding to prompt AI support (RQ3). 
\end{itemize}

Our paper concludes with a discussion on the importance of being able to measure how roles are enacted, the dangers of mislabeled roles creating impossible expectations on human supporters, and a need for new labels to describe new, emergent AI-native roles.
 
\section{Related Work}

\subsection{Role Theory and Role Conflict}

Role theory offers a vocabulary for how people organize behavior around social positions~\cite{biddle_recent_1986}. \citet{linton_study_1936} distinguished a \emph{status}, the position a person occupies, from a \emph{role}, the dynamic aspect of that status: what the person does by virtue of occupying it.
Later work separated the expectations attached to a position from the behavior actually performed in it~\cite{biddle_recent_1986}.
We call what a person is expected to do in a given position the \emph{prescribed role} (e.g., a teacher is expected to provide instruction), and what the person actually does the \emph{enacted role} (e.g., a teacher may also act as a parental figure). 
Common approaches to designing AI supporters, rely on persona injection which focus on defining a \emph{prescribed role}, but we lack methods to evaluate \emph{enacted role}.

Roles are not fixed traits. They are enacted over time and shift with the interaction partner and the conversational context. \citet{turner_role-taking_1962} framed role-taking as an ongoing process rather than conformity to a fixed script, with participants continually making and remaking their roles in relation to one another. When a person faces two or more incompatible expectations at once, the result is \emph{role conflict}, which studies have associated with stress, poorer job performance, and higher turnover~\cite{biddle_recent_1986}. AI role prompting prescribes a role once at the start but provides no turn-by-turn guidance. Thus, unlike human support, it is unclear how rigid or transient enacted roles in AI support are.

\subsection{AI in Support-Seeking} 
AI chatbot functions have expanded beyond task assistance into companionship, emotional support, and romantic relationships~\cite{manoli_shes_2025}. AI systems today are simultaneously addressed as therapist, coach, friend, and informant, and the role they enact may blend within a conversation or drift across it. People increasingly turn to conversational systems for emotional and psychological support because of their accessibility, anonymity, and perceived non-judgment~\cite{song_typing_2025}. Researchers have explored LLM support in various forms, from motivational interviewing coaches that use non-directive questioning to elicit change talk~\cite{jorke_gptcoach_2025,meyer_you_2024} to emotionally supportive coworkers for front-desk workers~\cite{das_swain_ai_2025}. Users report that AI coworkers help them avoid negative thinking and recenter their thoughts, but the same study also surfaces limitations of such support. For instance, an AI coworker cannot share lived experience, something that is indispensable to peer support~\cite{das_swain_ai_2025}. AI is increasingly cast in supportive roles whose relational obligations, such as sharing lived experience, it cannot truly fulfill.

The role an AI system occupies is not fixed at deployment.
People have long been known to treat computers as social actors, projecting onto them human-like qualities that might not have been programmed into the AI system~\cite{nass_computers_1994}. These qualities accrue through repeated interaction, sometimes leading to the user developing reliance or even emotional attachment, taking on romantic roles when that role was not prescribed ~\cite{pataranutaporn_my_2025}. AI systems that take on the role of a companion can also introduce a distinct class of harms, including relational transgression, harassment, and privacy violations~\cite{zhang_dark_2025}. Furthermore, vulnerable populations can be manipulated through LLM interactions~\cite{shen_hidden_2026}.  Harmful responses may go unnoticed due to unquestioning user trust~\cite{choi_examining_2025}. Users’ tendency to treat AI as a social actor can lead them to form attachments and cast AI into roles that was not prescribed, exposing them to dependence, relational harm, manipulation, and unnoticed harmful responses.
 
Role instability also originates in the model itself.
LLMs can also drift from their prescribed role~\cite{gonnermann-muller_maintaining_2026} and display sycophancy, a tendency to give socially desirable responses that causes further deviation~\cite{cheng_sycophantic_2026}. Over time, LLMs' behavior deviates from the prescribed personas over extended interactions~\cite{araujo_persistent_2026,ding_contextecho_2026}. \citeauthor{purohit_conditional_2026} show that LLM role adherence is \emph{conditional}: models maintain prescribed roles only when conditions remain stable, and deviate when context shifts~\cite{purohit_conditional_2026}. Unlike humans, LLMs do not experience role conflict as stress, but prompts and interaction contexts can still impose conflicting expectations, and models may produce behavior consistent with role conflict~\cite{zeng_who_2025,shin_roleconflictbench_2026}. Recent work shows that LLMs can internalize role conflict~\cite{zeng_who_2025}, and their decisions in role-conflict scenarios are shaped by social-role preferences~\cite{shin_roleconflictbench_2026}. This matters because a single AI application could be repurposed for multiple needs, which can lead to it attempting to perform a conflicting positions where it takes up multiple roles or deviates from the prescribed one. 

\subsection{Evaluating AI Behavior in Support-Seeking}  
The current research landscape is well aware of these challenges and dangers, and many studies and evaluation methods have been developed to detect when an AI's behavior does not align with its prescribed role or ethical guidelines. Some approaches focus on domain-specific contexts. For example, the Motivational Interviewing Treatment Integrity (MITI) coding system scores fidelity to motivational interviewing (MI) practice at the session level~\cite{moyers_motivational_2016} and has been used to evaluate LLM coaches using MI~\cite{jha_benchmarking_2026,oster_can_2026}. BOLT~\cite{chiu_computational_2024}, a computational framework for assessing LLM therapist behavior, finds that LLM responses often resemble low-quality therapy rather than high-quality therapy. These evaluation methods are invaluable for understanding how well an AI system performs a specific prescribed role. However, they do not specify how to describe, distinguish, or measure adherence across the multiple roles that an LLM may enact during support-seeking.

Our proposed approach, \AROMA aims to address these gaps in our understanding of AI support by looking at how we understand roles from a Role Theory perspective. 

\section{Analytical Framework}
\label{sec:methods} 
Existing evaluation methods assess agent behavior based on the values and goals they are committed to, these approaches do not account for general-purpose AI agents in multi-use settings and fail to provide domain-agnostic insights on AI behavior \cite{shin_roleconflictbench_2026, lai_rolecdebenchmarking_2026, song_role_2026}. This approach does not account for the multi-faceted and fluid nature of human role adaptation in a conversation. To improve our understanding of AI support in naturalistic settings, we focus on inspection of enacted roles across different domains. We propose a theory-grounded data-driven method to evaluate role dynamics, that we call \AROMA (A Role-Oriented Model of Analysis). \AROMA establishes what role an AI system actually performs, without knowing what role it was assigned. 

It provides a taxonomy grounded in established theory, applied across different datasets spanning peer support, motivational interviewing, and counselling. 
For each context, we establish human baselines  (RQ1), compare AI enactment against those baselines (RQ2), and test whether the resulting divergence can be reduced through steering (RQ3).
This section describes the shared framework; the method and results for each research question follow in Sections~\ref{sec:rq1}--\ref{sec:rq3}.


\subsection{Corpora Composition}
\label{sec:corpus}
To ensure our framework is representative of actual AI support behavior, we set out to compile a corpus of publicly available datasets of real \textit{support conversation}. Throughout the paper, \textbf{support conversation} will refer to any dialogue in which one party (the \emph{supporter}) intentionally assists another (the \emph{seeker}) in managing emotional distress, exploring a problem, or changing behavior. A \textbf{support role} is the pattern of behaviors a supporter enacts over the course of such a conversation: not the label they are given but what they observably do, turn by turn. Each corpus, that we compiled, included datasets from a fixed support context. For our study, we needed each corpus to contain at least one dataset where support conversations involve a human seeker conversing with human support (\hh)  \cite{wu_anno-mi_2022, malhotra_speaker_2022} and a human seeker conversing with an AI supporter (\hai). \cite{meyer_you_2024, yan_sharechat_2026} These different pairings of seeker and supporter, or as we call them \textbf{dyads}, were important for us to be able to learn how AI behavior might deviate from human behavior in the same setting. Additionally, we expanded any corpus with AI seeker and AI supporter dyads (\aiai) as these are commonly used in research for simulation and inform many of the AI support agents within the scope of our work ~\cite{chen_socialsim_2025, lee_cactus_2024}. Based on these criteria, we identified datasets representing three different contexts that span a spectrum of support settings:


\begin{table}[t]
  \caption{Corpus composition. The corpus comprises 99,872 coded segments across 7,515 conversations from 13 datasets in three languages (English, Chinese, and Japanese). SENSE-7 contains only a single dyad type and is excluded from regression models.}
  \label{tab:datasets}
\Description{Table listing 13 datasets across MI, Counselling, and Peer contexts with dyad types, languages, segment counts, and conversation counts. Total: 99,872 segments across 7,515 conversations.}
  \centering
  \small
  \begin{tabular}{@{} l l l r r >{\raggedright\arraybackslash}p{0.38\linewidth} @{}}
    \toprule
    Dataset & Dyad & Lang & Segments & Convs & Description \\
    \midrule
    \multicolumn{6}{@{}l}{\textit{Motivational Interviewing (3{,}991 segments, 317 conversations)}} \\
    AnnoMI          & \hh   & EN & 2{,}118  & 132   & Transcribed MI therapy sessions between trained clinicians and real clients, annotated for therapist behaviour and MI fidelity~\cite{wu_anno-mi_2022} \\
    MICha           & \hai  & EN & 1{,}873  & 185   & Crowdsourced conversations in which real users discussed behaviour change with a GPT-4-based MI chatbot; each turn rated for helpfulness~\cite{meyer_you_2024} \\
    \midrule
    \multicolumn{6}{@{}l}{\textit{Counselling (71{,}089 segments, 4{,}072 conversations)}} \\
    HOPE            & \hh   & EN & 2{,}929  & 158   & Transcribed counselling sessions from publicly available YouTube videos, annotated with speaker-level dialogue acts~\cite{malhotra_speaker_2022} \\
    KokoroChat      & \hh   & JA & 21{,}369 & 1{,}427 & Japanese counselling dialogues between trained counsellors and crowd-workers in structured role-play scenarios~\cite{qi_kokorochat_2025} \\
    PsyDial         & \hh   & ZH & 996      & 87    & Chinese counselling dialogues from professional counselling textbooks (D101 evaluation set); both counsellor and client are human-authored~\cite{qiu_psydial_2025} \\
    PsyDial-LLM     & \hai$^\dagger$ & ZH & 24{,}226 & 700   & Privacy-preserved counselling dialogues (D4) in which real human counsellor responses are retained and client utterances are LLM-reconstructed via the RMRR pipeline~\cite{qiu_psydial_2025} \\
    CACTUS          & \aiai & EN & 9{,}661  & 700   & LLM-generated cognitive-behavioural therapy dialogues between a simulated therapist and client~\cite{lee_cactus_2024} \\
    SimPsyDial      & \aiai & ZH & 11{,}908 & 1{,}000 & Fully LLM-generated counselling dialogues using a three-stage therapeutic model (exploration, comfort, action)~\cite{qiu_interactive_2024} \\
    \midrule
    \multicolumn{6}{@{}l}{\textit{Peer Support (23{,}863 segments, 2{,}867 conversations)}} \\
    ESConv          & \hh   & EN & 12{,}205 & 1{,}300 & Crowd-worker conversations in which a trained supporter helps a help-seeker through an emotional difficulty, annotated with eight support strategies~\cite{liuEmotionalSupportDialog2021} \\
    EmpDial         & \hh   & EN & 1{,}916  & 663   & Short empathetic exchanges in which a listener responds to a speaker sharing a personal emotional experience~\cite{rashkin_towards_2019} \\
    ShareChat       & \hai  & EN & 2{,}086  & 204   & Naturally occurring conversations between real users and commercial LLM chatbots, voluntarily shared on a public platform~\cite{yan_sharechat_2026} \\
    SocialSim       & \aiai & EN & 7{,}656  & 700   & Fully synthesised emotional support dialogues generated with explicit social-disclosure modelling between two LLM agents~\cite{chen_socialsim_2025} \\
    \midrule
    \multicolumn{6}{@{}l}{\textit{Excluded from regression (929 segments, 259 conversations)}} \\
    SENSE-7         & \hai  & EN & 929      & 259   & Conversations between real users and an empathetic AI companion designed for general emotional support~\cite{suhSENSE7TaxonomyDataset2025} \\
    \bottomrule
  \end{tabular}
  \vspace{0.3em}
  \par\noindent\footnotesize
  $^\dagger$In PsyDial-LLM the \emph{supporter} is human and the \emph{seeker} is LLM-reconstructed, the reverse of MICha and ShareChat where the supporter is AI.
\end{table}

\begin{itemize}
    \item \textbf{Motivational Interviewing (MI)} is a person-centred, goal-oriented method for enhancing intrinsic motivation by exploring and resolving ambivalence~\cite{miller_motivational_1991}. Four principles guide MI: expressing empathy, developing discrepancy, rolling with resistance, and supporting self-efficacy~\cite{miller_motivational_2002}. MI uses non-directive questioning as a method of nudging the seeker to articulate their own goals. The outcome of MI is increase in agency due to goal ownership. We selected 2 data sets for MI, AnnoMi \cite{wu_anno-mi_2022} (\hh; expert-annotated MI demonstration dialogues from YouTube videos) and MICha \cite{meyer_you_2024} (\hai;  GPT-4 conversations with human seekers, prompted with MI instructions).  
    \item \textbf{Counselling}, in this work, practices reflective listening and open-ended questioning to facilitate client exploration and insight. Restatements and reflections strengthen the working alliance~\cite{rautalinko_reflective_2013}, helping clients articulate and examine their experiences. Counselling uses a range of tools, including Cognitive Behavior Therapy. For Counselling we have CACTUS \cite{lee_cactus_2024} ( \aiai LLM-generated cognitive-behavioural therapy dialogues between a simulated therapist and client), HOPE \cite{malhotra_speaker_2022} ( \hh Transcribed counselling sessions from publicly available YouTube videos), KokoroChat \cite{qi_kokorochat_2025} (\hh Japanese counselling dialogues between trained counsellors and crowd-workers in structured role-play), PsyDial \cite{qiu_psydial_2025} (which consisted of 2 types of data, \hh; human-authored textbook counselling dialogues, and \hai; LLM generated datasets which had LLM-reconstructed client conversation for data privacy), SimPsyDial \cite{qiu_interactive_2024} (\aiai fully LLM-generated psychotherapy dialogues) 
    \item \textbf{Peer support} is ``a system of giving and receiving help founded on key principles of respect, shared responsibility, and mutual agreement of what is helpful''~\cite{mead_peer_2001}. It differs from professional care in that supporters draw on shared lived experience to build reciprocal rapport. Expressing empathy through disclosing one's lived experiences helps the seeker feel heard and not alone. For Peer datasets, we selected ESConv \cite{liuEmotionalSupportDialog2021} (\hh; crowd-worker conversations collected through role-play), EmpDial \cite{rashkin_towards_2019} (\hh; Short empathetic exchanges between paired MTurk crowd-workers),  ShareChat \cite{yan_sharechat_2026} Dataset  (human-seeker conversations with AI chatbots interactions, filtered to personal issues), SocialSim \cite{chen_socialsim_2025} (\aiai  synthetic data that uses ESConv datasets as few shot examples) Sense-7 (\hai empathetic AI companion designed for general emotional support). 
\end{itemize}

Our full corpus comprises 99,872 coded segments across 7,515 conversations from 13 datasets in three languages (English, Chinese, and Japanese) can be found in \autoref{tab:datasets}.


\subsection{Theory-Guided Bottom-Up Role Annotation}

The rich corpora of data we established helped us develop a taxonomy for support roles from the bottom-up. To encode different types of roles, we referred to the literature on social role theory to derive fundamental definitions of roles. 

\textbf{Dimensions of a Support Role}: In general, we found that the literature does not describe canonical roles such as ``friend'', ``mentor'' or ``parent''. Instead, prior work in this domain defines roles through different dimensions. Our framework predominantly relies on two such dimensions. The first is \textit{Orientation}, which was developed using Hill's helping-skills model~\cite{hill_helping_2010}, which identifies 12 verbal response modes that supporters use, including asking, reflecting, restating, and advising. These responses are further organised along the temporal evolution of the support conversation (exploration, insight, action). This model was particularly appropriate for our framework because it accounts for a variety of transitioning response modes within a single conversation. The second is \textit{Support Type} which was based on Cutrona and Russell's optimal-matching theory~\cite{cutrona_type_1990}, which distinguishes five support types along a nurturant---action-facilitating axis, providing the vocabulary for \emph{what kind} of support a given response provides.

\textbf{Initial Coding Process to Refine Dimensions}: We operationally defined a conversational \textit{segment} as a sequence of three turns that begins with a seeker's message, a supporters response, and then the seeker's next message. Two co-authors independently coded the same segments for the \textit{Orientation} and \textit{Support Type}. This initial pool was a stratified sample of $57$ segments from our corpora. For the first phase of coding we relied on the definitions of from Hill's helping-skills~\cite{hill_helping_2010} and Cutrona and Russell~\cite{cutrona_type_1990}. The coders inspected each segment and assigned labels for each dimensions separately. For \textit{Orientation}, in particular, the coders found that a supporter might demonstrate multiple different helping skills within the same segment. In fact, human supporters had exhibited simpler turns, \cite{hill_helping_2010} {``So have your feelings about marijuana changed since we last talked?''}\cite{malhotra_speaker_2022}. However, a typical AI turn from our counselling data below\cite{lee_cactus_2024}  illustrates a bundling reflection, validation, and a question. The seeker typically responds to only the dominant skill, answering the question, for instance, and the preceding reflection goes unacknowledged:

\begin{quote}
  \textit{``That must have felt very overwhelming.''} [reflection] \\
  \textit{``It's interesting that despite there being no direct negative feedback, you still felt that judgment deeply.''} [acknowledgement] \\
  \textit{``What do you think might be prompting these intense feelings of judgment and fear?''} [question]
\end{quote} 

Another challenge with direct adoption of the twelve Hill's skills as twelve labels for \textit{Orientation} was that some skills were fundamentally absent from AI-generated dialogues. AI has no lived experience to draw from, thus cannot share their own lived experiences to build empathy \cite{plaza-del-arco_no_2025, lu_assistant_2026} .  In light of these nuances in support skills, we decided to define \textit{Orientation} labels as a composite of skills. This approach of abstraction was inspired by \citeauthor{mccarthy_multitheoretical_2009}'s MULTI framework~\cite{mccarthy_multitheoretical_2009}, which grouped 60 observable actions of therapists into eight orientations to capture their actions.  
In the same vein, we condensed the Hill's skills~\cite{hill_helping_2010} into four categories: 
 \textbf{Asking} ( Open/closed questions ), \textbf{Repeating Back} ( Restatement, reflection, interpretation ), \textbf{Acknowledgement} (reassurance, personal experience) , and \textbf{Advising} (guidance, information, challenge) \cite{hill_helping_2010}. This abstraction allowed us to descriptively annotate for \textbf{orientation} to describe \emph{how} the supporter responds to the seeker. \autoref{tab:orientations} lists the four orientation.
Similarly, for Cutrona and Russell's optimal matching model~\cite{cutrona_type_1990} we found support types such as ``tangible'' to rarely occur in our dataset, so instead we coded for the macro-categories they defined two \emph{support types}: \textbf{Nurturant} (emotional and esteem-focused) and \textbf{Action-Facilitating} (tangible and informational).



\begin{table}[t]
  \centering
  \caption{The four Orientations. (Hill, 2009) provide turn-level evidence; Orientation captures the dominant function of the turn as a whole.}
  \label{tab:orientations}
  \Description{Table defining four orientations: Asking opens new directions, Repeating Back mirrors seeker content, Acknowledgement validates without processing specifics, Advising delivers new content.}
  \small
  \begin{tabular}{@{} l p{0.35\textwidth} p{0.45\textwidth} @{}}
  \hline
  \textbf{Orientation} & \textbf{Supporter's action} & \textbf{D1 spans} \\
  \hline
  Asking & Asks questions & Open/closed questions \\
  \hline
  Repeating Back & Echoes, restates, or names seeker's expression & Restatement, reflection, interpretation \\
  \hline
  Acknowledgement & Validates; normalises; builds esteem & Approval, reassurance, personal experience \\
  \hline
  Advising & Gives external guidance, info, or plans & Direct guidance, information, challenge \\
  \hline
  \end{tabular}
\end{table}






 


\textbf{Iterative Manual Coding:} After refining our definition of role dimensions, the two coders iteratively annotated segments in batches. We re-coded our pilot batch ($n = 57$) to improve the definitions of the different categories of \textit{Orientation} and \textit{Support Type}. The principal confusions were advising vs.\ asking and acknowledgement vs.\ advising. The coders discussed their rationale to improve the descriptions further and clarifying the boundary between categories within dimensions. We further iterated on $496$ additional stratified samples of segments and broken down into four waves of coding. Finally, we evaluated our codebook descriptions and  inter-rater reliability on 100 held-out segments stratified across 13 datasets. Orientation achieved substantial agreement (Cohen's $\kappa = 0.713$, 95\% CI $[0.607, 0.814]$, 79\% observed agreement). For support type, observed agreement was 81\%; because the distribution is heavily skewed toward Nurturant ($\sim$83\%), we report Gwet's AC1 $= 0.740$, 95\% CI $[0.599, 0.856]$, rather than Cohen's $\kappa$, which is deflated by high base rates. The final descriptions of our codes for both dimensions are described in \autoref{tab:decision-rules}.

\begin{table}[ht]
  \caption{Coding rules for Orientation and Support Type. Orientation is determined by the dominant function of the supporter's turn. Support Type is coded independently. The full decision procedure with worked examples is in the supplementary material.}
  \label{tab:decision-rules}
  \Description{Table specifying coding rules for five orientations and two support types. Each row defines the decision procedure for assigning one label.}
 
  \centering
  \begin{tabular}{@{} l p{0.70\textwidth} @{}}
    \toprule
    Label & What the supporter is doing \\
    \midrule
    \multicolumn{2}{@{}l}{\textsc{Orientation} --- \textit{how the supporter responds to what the seeker just said}} \\[4pt]
    Asking          & Asks a question that opens something new: the seeker's feelings, experiences, or history. Questions that recommend a specific action (``Why don't you try X?'') count as Advising. \\[2pt]
    Repeating Back  & Gives back what the seeker said: echoing their words, naming their feelings, or rephrasing their meaning. \emph{Test}: swap the seeker's details for unrelated ones. If the turn no longer makes sense, it is Repeating Back. \\[2pt]
    Acknowledgement & Validates or supports the seeker without working with their specific content. ``I hear you'' and ``That takes courage'' work the same way regardless of what the seeker said. Includes sharing personal experience as solidarity. \\[2pt]
    Advising        & Gives the seeker something new: advice, a plan, information, or a reframe that pushes toward change. \emph{Test}: remove the advice. If nothing meaningful remains, it is Advising. \\[2pt]
    Other           & Not a helping skill (session logistics, AI errors, greetings). \\
    \midrule
    \multicolumn{2}{@{}l}{\textsc{Support Type} --- \textit{what kind of support the turn provides}} \\[4pt]
    Nurturant              & Attends to feelings, affirms worth, or reminds of social ties. \\[2pt]
    Action-Facilitating    & Delivers knowledge, options, or concrete resources. \\
    \bottomrule
  \end{tabular}

  \vspace{0.5em}
  \par\noindent\footnotesize
  \emph{Mixed turns.} When a turn bundles multiple moves (common in AI responses), the classifier tags each move, then assigns the orientation of whichever function carries the turn. When the turn itself is ambiguous, the seeker's next response serves as a tie-breaker.
\end{table}

\textbf{Using LLM-as-a-judge to Code the Corpora:} To scale annotation to the full corpus, we developed an LLM-based classifier for each dimension  (see Appendix~\ref{app:d2-classifier-prompt} and Appendix~\ref{app:d3-classifier-prompt} ) using DeekSeek-V3 \cite{deepseek-ai_deepseek-v3_2025} that implements the decision procedure in Table~\ref{tab:decision-rules} as a structured prompt. We iterated the prompt against codes with human consensus ($n = 160$ of the same segments for human coding) and audited gold labels against classifier disagreements across the five iterative coding sessions ($n=300$ of the human coding segments). Finally, we validated the final LLM prompt against $79$ of the 100 held-out segments on which both human coders agreed. The classifier attained $\kappa = 0.734$, 95\% CI $[0.610, 0.844]$ (81\% agreement) on orientation and AC1 $= 0.889$, 95\% CI $[0.792, 0.959]$ (91\% agreement) on support type, placing it within the range of human inter-rater reliability. We applied this classifier to annotate the full corpus.

\begin{table}[t]
  \caption{The eight Role States (Orientation $\times$ Support Type). Each Role State combines a conversational orientation (how the supporter moves the exchange—ask, reflect-back, acknowledge, or advise) with a support type (nurturant vs.\ action-facilitating). Examples are from human conversations. A \emph{role signature} is the distribution of these states across a conversation; we use signature divergence to measure role misalignment.}
  \label{tab:role-states}
    \Description{Table showing eight role states formed by crossing four orientations with two support types. Each cell contains a role name and example from a human conversation.}

  \centering
  \begin{tabular}{@{} l l >{\raggedright\arraybackslash}p{0.75\linewidth} @{}}
    \toprule
    Role State &    & Example \\
    \midrule
    \textbf{Exploring}    & Ask $\times$ Nur & \textit{``What makes you say `I have to take that path'?''} \\
    \textbf{Suggesting}   & Ask $\times$ AF  & \textit{``Is there an HR department you can take it to?''} \\
    \textbf{Reflecting}   & RB $\times$ Nur  & \textit{``It doesn't quite fit with how you see your own lifestyle.''} \\
    \textbf{Formulating}  & RB $\times$ AF   & \textit{``So it does appear you have phobic anxiety symptoms.''} \\
    \textbf{Affirming}    & Ack $\times$ Nur & \textit{``You work extremely hard and deserve that raise.''} \\
    \textbf{Relating}     & Ack $\times$ AF  & \textit{``I had same issue with my sleep pattern and I did few things such as stretching exercise one hour before sleep.''} \\
    \textbf{Reframing}    & Adv $\times$ Nur & \textit{``Think of it as getting a new beard, rather than losing the old friend.''} \\
    \textbf{Directing}    & Adv $\times$ AF  & \textit{``I would go above your supervisor since nothing has been done.''} \\
    \bottomrule
  \end{tabular}
\end{table}

\subsection{AROMA's Role Taxonomy}

With LLM-assistance, we coded \textit{support type} and \textit{orientation} on our support seeking corpora. Crossing the two dimensions yields eight \rolestates, such as \RoleExploring (\AskNur ) and \RoleReframing (\AdvNur). Each capturing both the manner of delivery and the kind of support provided. \autoref{tab:role-states} lists out the 8 \rolestates.

\begin{figure}[t]
  \centering
  \includegraphics[width=\linewidth]{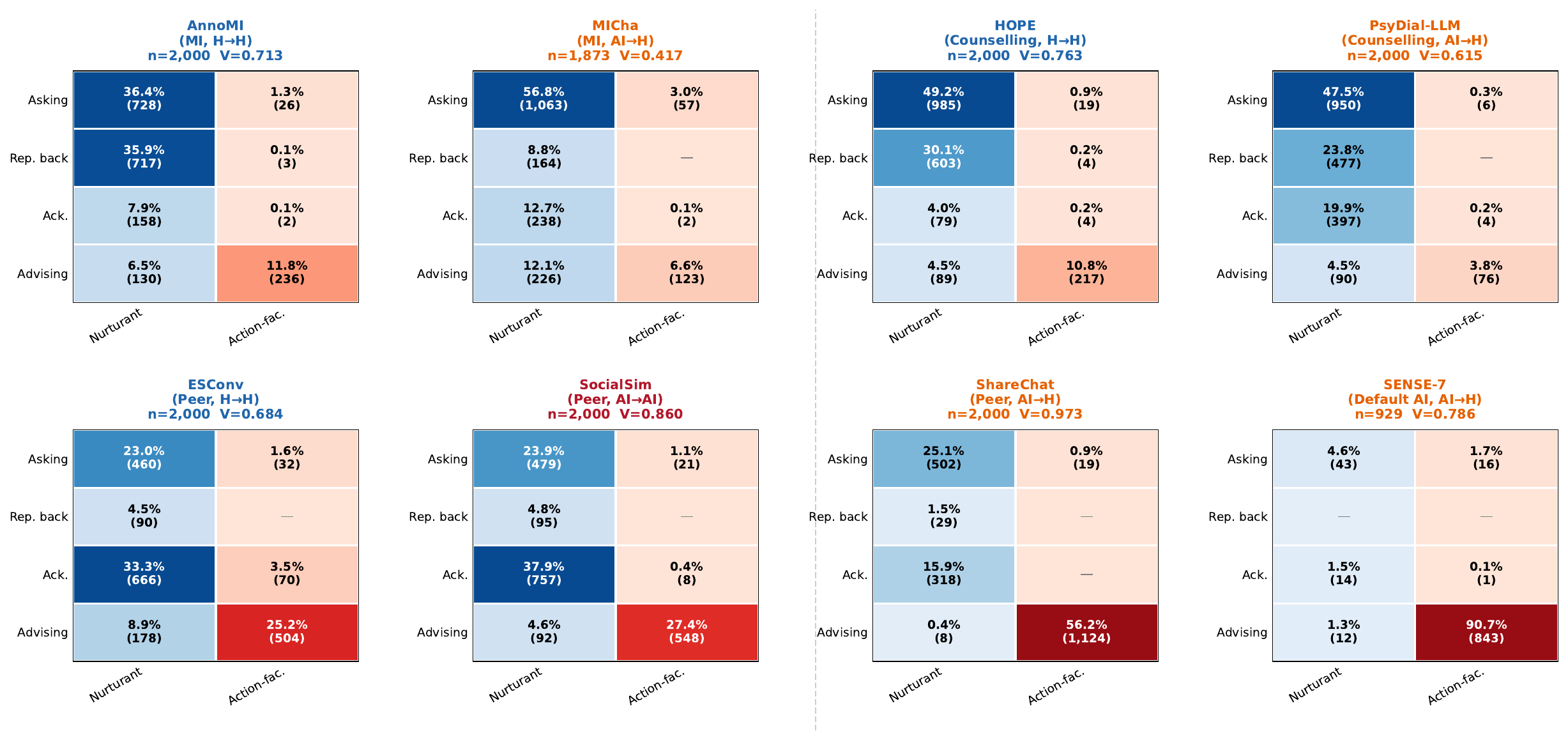}
  \caption{Orientation $\times$ support type distributions for each dataset in the corpus. Cell shading indicates the proportion of turns falling into each \rolestate. Cram\'{e}r's~$V$ quantifies how tightly orientation and support type are coupled within each dataset: higher values indicate that knowing one dimension largely determines the other.}
  \label{fig:coupling-grid}
  \Description{Grid of heatmaps showing orientation-by-support-type distributions for 13 datasets. Each cell's shading indicates the proportion of turns in that role state. Cramer's V values range from 0.42 (MICha, loosely coupled) to 0.97 (ShareChat, tightly coupled).}
\end{figure}

Although these dimensions are theoretically independent, we found the two role dimensions, to be coupled with each other at different degrees across datasets (~\autoref{fig:coupling-grid}). We quantified this coupling with Cram\'{e}r's~$V$, a measure of association strength between two categorical variables (0 = independent, 1 = perfectly associated). When coupling is high, knowing the orientation largely determines the support type; when coupling is low, the same orientation appears with both nurturant and action-facilitating support. Across the corpus, $V$ ranges from 0.42 (MICha) to 0.97 (ShareChat). AI systems occupy both extremes: MICha ($V = 0.42$) spreads across many orientation--support-type combinations, consistent with its system prompt, which discourages advice-giving. By comparison, in ShareChat ($V = 0.97$) if a supporter is enacting the asking orientation, then it is almost always with a nurturant support type (\RoleExploring) and advising orientation is always with action-facilitating. Having said that, we found that human datasets also show some degree of coupling between dimensions ($V = 0.60$--$0.75$), suggesting a flexible but structured relationship between orientation and support type. Moreover, two \rolestates, \RoleFormulating and \RoleRelating, occur in fewer than 1\% of our segments and were excluded from future analyses. We address our research questions through the remaining six \rolestates.

AROMA's taxonomy provides us with an operational definition of social roles that can be applied to any support seeking setting. Through AROMA we answer the key research questions of our study to shed light on the structural profiles of supporters in  different contexts (RQ1), the distinct attributes of AI supporters in comparison to humans (RQ2), and the effectiveness of dynamic role assignment in aligning AI support with that of humans (RQ3).

\section{RQ1: What are the fundamental features and structures of supporters' roles in different support-seeking context?}
\label{sec:rq1}
We know from role theory that distinct social context produces distinct patterns of conduct~\cite{biddle_recent_1986}. We developed AROMA to be able to provide a unified vocabulary of social roles across different support contexts. Therefore, to answer RQ1, we set out to analyze if the role dimensions described by AROMA can uniquely distinguish the support characteristics of different contexts (motivational interviewing, counseling and peer support). 


\subsection{Method}

We fitted a multinomial logistic regression to estimate the effect of context on \rolestate ($\text{role\_state} \;\sim\; \text{context}$). Here context was a categorical variable , we label MI, Peer Support, and Counselling as the three levels of context shared by both human and AI supporters. Similarly, for \rolestate, \RoleExploring was chosen as the reference level.
We restrict this analysis to \hh conversations only.
The sample for this analyses comprises 2,000 MI segments (132 conversations), 4,996 Counselling segments (977 conversations), and 3,916 Peer segments (1,702 conversations).
The regression used  counselling as a reference level to measure the odds ratios(OR), which we compared the other context categories against . 
Since segments are nested within conversations; to account for multinomial logistic regression's assumption of independent observations, we correct the model with cluster-robust standard errors from conversation-level bootstrapping (500 resamples; all converged).


\subsection{Results}



\begin{table}[t]
  \centering
  \begin{minipage}{0.48\textwidth}
    \centering
    \includegraphics[width=\linewidth]{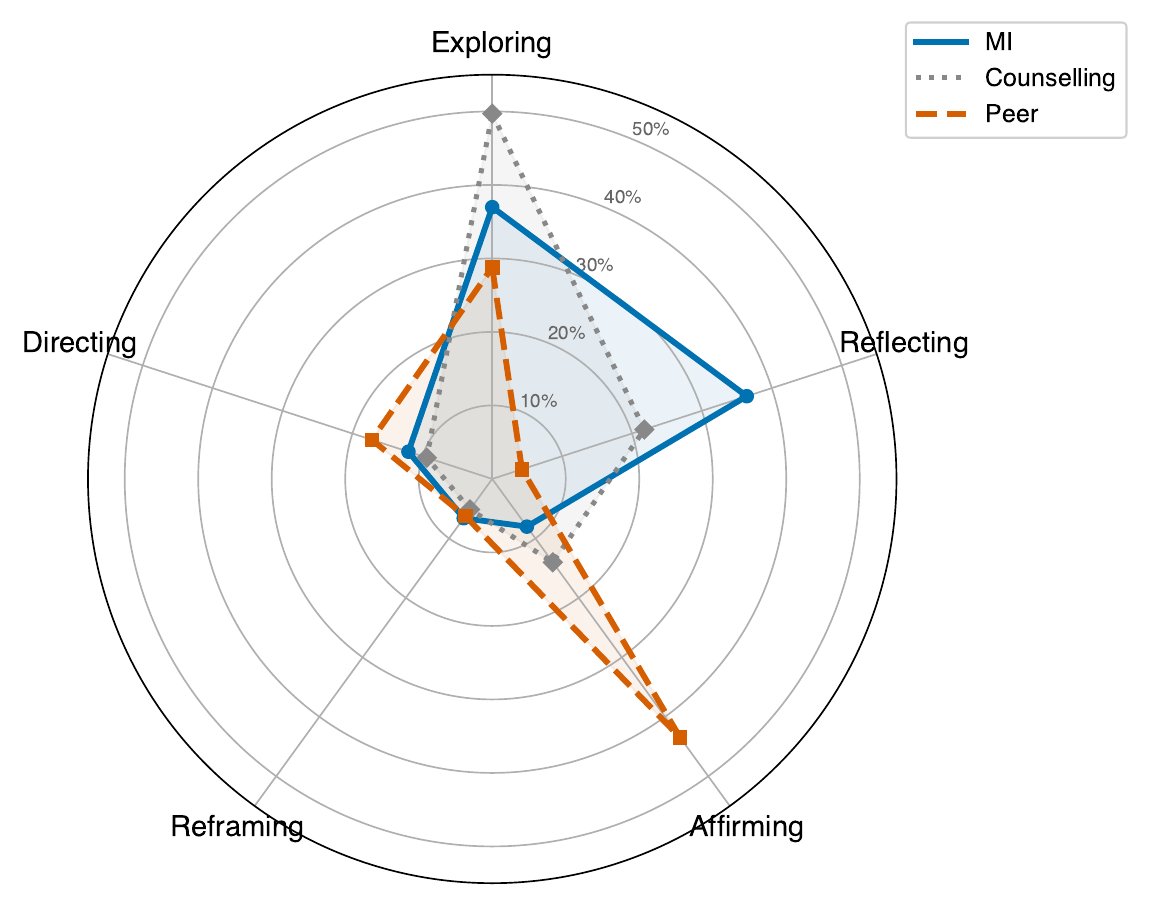}
    \captionof{figure}{Role-state profiles for each care context (human-only conversations).}
    \label{fig:context-radar}
      \Description{Radar chart showing role-state proportions for MI, Counselling, and Peer contexts. MI concentrates on Exploring and Reflecting; Peer concentrates on Affirming and Directing; Counselling distributes more evenly.}
  
  \end{minipage}
  \hfill
  \begin{minipage}{0.48\textwidth}
    \centering
    \small
    \captionof{table}{Effect of care context on role state, human-only conversations ($N = 10{,}912$ turns).
    Reference role: Ask$\times$Nur; reference context: Counselling. Model: \code{role $\sim$ context} (LR~$\chi^2 = 2387.7$, $\text{df} = 14$, $p < .001$; McFadden $R^2 = 0.071$)}
    \label{tab:rq1-predictions}
     \Description{Table of odds ratios for five role states comparing MI and Peer to Counselling. Peer Affirming OR equals 5.35; MI Reflecting OR equals 2.25. Reference role is Exploring.}
   
    \begin{tabular}{l | l l}
      \toprule
      Role State
      & MI vs.\ Counselling
      & Peer vs.\ Counselling \\
      \midrule
      Affirming
      & 0.77 [0.56, 1.06]
      & 5.35*** [4.60, 6.23] \\
      Directing
      & 1.72*** [1.28, 2.32]
      & 3.18*** [2.65, 3.81] \\
      Reframing
      & 1.72*** [1.30, 2.28]
      & 2.09*** [1.70, 2.58] \\
      Suggesting
      & 1.86* [1.07, 3.24]
      & 2.45*** [1.58, 3.81] \\
      Reflecting
      & 2.25*** [1.80, 2.82]
      & 0.34*** [0.28, 0.41] \\
      \bottomrule
    \end{tabular}
  \end{minipage}
\end{table}

Our regression revealed that each context was associated with distinct \rolestate profile (\autoref{tab:rq1-predictions}). Even though supporters in each context was where characterized by the same underlying dimensions (\textit{orientation} and \textit{support type}), the prominence of certain combinations of \rolestates differed between contexts. The regression model provides statistical evidence for differences that were only descriptively apparent in ~\autoref{fig:context-radar}. 


\textbf{Supporters in counselling rely on explorating:} Prior work shows that counsellors aim to build trusting relationships with their clients by using open-ended questions and reflective listening~\cite{rautalinko_reflective_2013}. Using AROMA, we found that human supporters frequently demonstrated \RoleExploring, which accounts for approximately half of all turns (49.1\%). Additionally, \RoleReflecting (21.5\%) and \RoleAffirming (13.9\%) emerged as other notable roles. This exploration-heavy distribution aligns with what is known from counselling research. Research shows that counsellors often employ non-directive skills~\cite{rautalinko_reflective_2013}. On closer inspection of our roles we find that the dominant roles for these types of supporters tend to have the Nurturant support type, as opposed to Action-Facilitating, which tends to  provide support through actionable suggestions. 

\textbf{Supporters in MI show increased reflecting and directing:} In MI, the aim of supporters is to enable behavior change in seekers through reflective listening and questions~\cite{miller_motivational_2002}. The mechanisms are similar to what counsellors apply, but the intent and outcomes are different. Supporters often repeat what a seeker is saying to help them perceive the gap between their current behavior and their personal goals, to intrinsically motivate change.
AROMA's analyses shows that  \RoleReflecting accounts for 35.9\% of MI turns, nearly matching \RoleExploring (36.4\%). Compared to counselling, MI shows elevated \RoleReflecting (OR = 2.25 [1.80, 2.82]), along with higher \RoleReframing (OR = 1.72 [1.30, 2.28]) and \RoleDirecting (OR = 1.72 [1.28, 2.32]). \RoleAffirming does not differ significantly (OR = 0.77 [0.56, 1.06]) from counselling, which fits our understanding of the two different contexts. This style of reflection emphasizes  MI's core principle of mirroring the seeker's words towards change. 

\textbf{Supporters in peer settings show increased affirming:} Finally, we study the roles among peer supporters. In this context, supporters tend to draw on shared lived experience to validate the seekers experience and advise accordingly~\cite{mead_peer_2001}. AROMA found that 
\RoleAffirming was the most frequent state (41.8\%), followed by \RoleExploring (27.7\%) and \RoleDirecting (16.5\%). 
Relative to counselling, peer support shows increases in \RoleAffirming  (OR = 5.35 [4.60, 6.23]) and \RoleDirecting (OR = 3.18 [2.65, 3.81]). We also know that peers rarely employ the formal reflective techniques central to professional practice \cite{mead_peer_2001}. In accordance to this, we also observe that \RoleReflecting was suppressed (OR = 0.34 [0.28, 0.41]) compared to counselling.  This pattern is representative of the peer-support norm of validating first, then offering experience-based guidance.

These insights show us that human support across context can be described using the common taxonomy of AROMA's \rolestates. In this way, AROMA provides a parsimonious and comparable representation of support roles. 


\section{RQ2: How do AI support roles differ from that of human actors?}
\label{sec:rq2}

Role theory holds that social roles are enacted dynamically, shifting in response to the interaction partner and the conversational context~\cite{biddle_recent_1986}. LLMs, by contrast, are given a single prescribed role via a system prompt \cite{tseng_two_2024} and have no social identity to maintain or adhere to beyond that prescription \cite{plaza-del-arco_no_2025, lu_assistant_2026} . Therefore, we hypothesize that AI supporters will exhibit less variability in their role states, producing a narrower and more rigid profile of roles. We test this hypothesis through two analyses to compare differences between human and AI actors: first, we model the static distribution of role states as a function of supporter and seeker identity; second, we analyze role states as a dynamic transitioning property of different \hh, \hai, and \aiai conversations.

\subsection{Variation in structure of human and AI support roles}

\begin{equation}
  \text{role\_state}_t \;\sim\; \text{supporter\_type} + \text{seeker\_type} + \text{context} + \text{length} + \text{position}
   \label{eq:rq2-seeker-support-model}
\end{equation}

\subsubsection{Method} We build upon our multinomial logistic regression from RQ1's analyses (\autoref{eq:rq2-seeker-support-model} to estimate how \rolestate is associated with supporter type (human or AI), seeker type (human or AI). Our model incorporates support context, turn length, and conversational position~\footnote{This is analogous to turn count, with a higher position implying a higher turn count.}
We use the stratified sample of 22{,}785 segments from 6{,}025 conversations across 12 datasets described in Section~\autoref{sec:corpus}. Again, we cap each dataset at 2{,}000 randomly selected segments to prevent large corpora from dominating coefficient estimates. A sensitivity analysis varying the dataset cap (1{,}000 and 3{,}000) confirmed that all odds ratios were stable.
The reference role is \RoleExploring (\AskNur), the reference identity is Human for both supporter and seeker, and the reference context is Counselling. 
 
The regression fits the data well (McFadden $R^2 = 0.169$; LR~$\chi^2 = 11{,}504.9$, $\text{df} = 42$, $p < .001$).

\begin{table}[t]
  \centering
  \caption{Odds ratios for AI supporter and AI seeker on each role state, relative to human, from the multinomial logistic regression in Equation~\ref{eq:rq2-seeker-support-model}.}
  \label{tab:rq2-seeker-supporter}
   \Description{Table of odds ratios comparing AI to human supporters and seekers across five role states. AI supporters show Reflecting OR of 0.09 and Affirming OR of 0.33. AI seekers show modest increases in Directing and Suggesting.}
  \begin{tabular}{l rl rl}
    \toprule
    & \multicolumn{2}{c}{Supporter (AI vs.\ Human)} & \multicolumn{2}{c}{Seeker (AI vs.\ Human)} \\
    \cmidrule(lr){2-3} \cmidrule(lr){4-5}
    Role State & OR & 95\% CI & OR & 95\% CI \\
    \midrule
    \RoleAffirming   & 0.33*** & [0.27, 0.39] & 1.44*** & [1.17, 1.77] \\
    \RoleDirecting   & 0.55*** & [0.45, 0.67] & 1.71*** & [1.37, 2.14] \\
    \RoleReframing   & 0.64*** & [0.53, 0.77] & 0.99 & [0.80, 1.22] \\
    \RoleSuggesting  & 1.22 & [0.93, 1.62] & 1.51** & [1.12, 2.04] \\
    \RoleReflecting  & 0.09*** & [0.08, 0.11] & 1.24* & [1.04, 1.49] \\
    \bottomrule
  \end{tabular}
\end{table}


\subsubsection{Results}
~\autoref{tab:rq2-seeker-supporter} reports odds ratios for AI supporter and AI seeker, each relative to human, holding the remaining predictors constant.

\textbf{AI supporters default to exploring:} Relative to human supporters, AI supporters concentrate their behavior towards \RoleExploring and suppress every other role state (~\autoref{tab:rq2-seeker-supporter}). Moreover,  \RoleSuggesting remains comparable to human rates (OR = 1.22 [0.93, 1.62]). Both of these roles involve the Asking orientation, which falls in line with standard LLM behavior to continue conversations through questions and task-driven suggestions \cite{lu_assistant_2026}. Compared to human supporters, the suppression is most severe for \RoleReflecting (OR = 0.09 [0.08, 0.11], $p < .001$): AI supporters mirror the seeker's content in fewer than one-tenth the turns that human supporters do. Similarly, \RoleAffirming is the next most suppressed (OR = 0.33 [0.27, 0.39], $p < .001$). Furthermore, we performed a Chi-square test to check the association between type of supporter and \rolestate to verify this result. The test corroborated this overall pattern ($\chi^2(7) = 2{,}180.4$, Cram\'{e}r's $V = 0.309$, $p < .001$), with the largest standardised residuals clustering on \RoleReflecting in the human direction. \RoleReflecting and \RoleAffirming, require the supporter to attend to the seeker's specific words and emotional state, then produce a response grounded in that content. Research indicates that LLMs do not organically extract the user's affective state and tend to  overweigh the instructional information of the message \cite{sharma_humanai_2023}. The consequence of these roles diminishing is that support seekers are less likely to build rapport or perceive support to be a collaborative meaning making activity, the way it can be with human supporters.

\textbf{AI seekers shift conversations modestly toward action:} Although roles are enacted by supporters, the seeker and their behavior are critical to how a supporter reacts \cite{stiles_responsiveness_1998}. The seeker's identity also shapes the role distribution, though the effects are smaller than the supporter effects. In comparison to humans, an AI-seeker shows increased \RoleDirecting (OR=1.71 [1.37, 2.14]), \RoleSuggesting (OR=1.51 [1.12, 2.04]), and \RoleAffirming (OR=1.44 [1.17, 1.77]). This seeker effect was largely observed in counselling context ($V = 0.229$), where AI-generated seekers (primarily used in simulated datasets) may present problems with less affective ambiguity, leading to more directive responses and fewer openings for reflective ones. Note, however, the probability of an AI seeker increasing the odds of certain role (e.g., an OR=1.24 implies the odds of AI supporters presenting \RoleReflecting traits is only 1.24$\times$ that of a human supporter) are much lower than an AI supporter reducing the odds of a role (e.g., an OR=0.09 implies the odds of  AI supporters presenting \RoleReflecting traits is almost 10$\times$ less likely than humans).  The modest size of these effects suggests that the supporter's identity is the primary driver of structural divergence, not the seeker's.

\subsection{AI and humans transition through roles differently}

\subsubsection{Method}  
Role theory has always described social roles are transient aspects of support setting \cite{turner_role-taking_1962}. This literature also informed us to define AROMA's roles as ``states'', not ``traits'', because of their fluid nature. To reemphasize all the \rolestate we have discussed were annotated at segment-level (or a sequence of seeker-supporter-seeker). To characterize this fluidity we investigate the following: 

\begin{enumerate}
    \item \textbf{Stability of \rolestates}: To measure how frequently different supporters switch (or retain) roles we simply compared the \rolestate between consecutive conversational segments.
    \item \textbf{Predicting next \rolestate}: To determine whether a past role state is associated with next role state, we extended the RQ2 regression (\autoref{eq:rq2-seeker-support-model}) by adding the previous turn's role state (where t = turn and t-1 = the prior turn)  as a variable (\autoref{eq:rq3-lagged-model}).
\end{enumerate}


\begin{equation}
  \text{role\_state}_t \;\sim\; \text{supporter\_type} + \text{seeker\_type} + \text{context} + \text{length} + \text{position} + \text{role\_state}_{t-1}
  \label{eq:rq3-lagged-model}
\end{equation}

\begin{figure}[t]
    \centering
    \includegraphics[width=\linewidth]{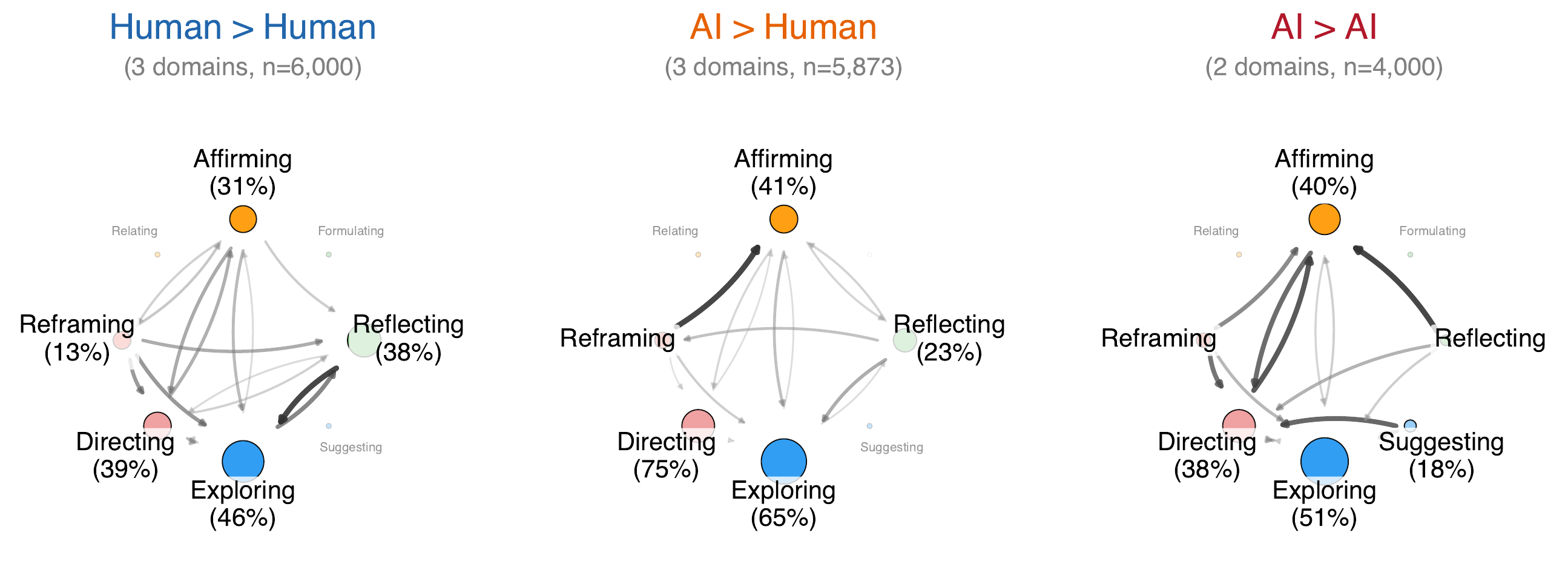}
    \caption{Role transition networks by conversation type (\hh, \hai, and \aiai). Node size reflects role frequency; edge width reflects conditional transition probability (edges below 8\% omitted).}
    \label{fig:transitions-matched}
      \Description{Three network diagrams showing role-state transitions for Human-Human, AI-Human, and AI-AI conversations. Node size reflects role frequency; edge width reflects transition probability. Human-Human shows diverse connections; AI conversations concentrate in Exploring with fewer transitions.}

\end{figure}

\subsubsection{Results} ~\autoref{fig:transitions-matched} illustrates the transition networks for each conversation type. 
\label{sec:results-rq2-transition}
\textbf{AI supporters are much less likely to switch roles:} \hh conversations switch roles on 62\% of consecutive turn pairs, compared with 36\% for \hai and 52\% for \aiai. 
\hh conversations distribute activity among \RoleExploring, \RoleReflecting, \RoleAffirming, and \RoleDirecting, with diverse transitions flowing among them. By contrast, AI conversations repeat the few \rolestates across consecutive turns more often: \RoleExploring is followed by another \RoleExploring in 71.5\% conversations of \hai and 57.9\% in \aiai (compared to  45.8\% of the time in \hh). AI conversations occupy fewer roles and stay in them longer, producing extended stretches of the same behaviour rather than the turn-by-turn adaptation characteristic of human exchanges.


\textbf{\rolestates indicate future \rolestates:} We fitted our model on $N = 16{,}760$ consecutive segment pairs~\footnote{Each conversation's first turn is excluded because it has no predecessor}. By including \(\text{role}_{t-1}\) to our original model (\autoref{eq:rq2-seeker-support-model} $\rightarrow$ \autoref{eq:rq3-lagged-model}) we find a significant improvement model fit: McFadden $R^2$ rises from 0.174 to 0.201 ($\Delta R^2 = 0.027$; LR~$\chi^2 = 1{,}401.8$, $\text{df} = 49$, $p < .001$). This insight ascertains that \rolestates do not emerge randomly but are indicated by the previous \rolestates.




%
\section{RQ3: How can AI supporters be designed for better role alignment?}
\label{sec:rq3}

From RQ2 we learned that AI support varies in how roles emerge and how the supporter switches between roles. Particularly, AI conversations are less dynamic. This limitation of existing AI agents begs us to question how it can be improved. Using AROMA, we propose to reduce the misalignment between human and AI supporters using a novel prompting approach.

\subsection{Method}
\begin{wrapfigure}{r}{0.35\textwidth}
\centering
\scalebox{0.5}{%
\begin{tikzpicture}[
    node distance=0.9cm and 2.2cm,
    every node/.style={align=center, font=\sffamily\small},
    box/.style={
        draw=black!50, rectangle, rounded corners=2pt,
        minimum width=3.8cm, minimum height=1.0cm,
        fill=white, inner sep=8pt
    },
    arrow/.style={-latex, draw=black!30, line width=0.5pt},
    feedback/.style={-latex, draw=black!70, line width=0.7pt, dashed}
]
\node[box] (step1) {
    \textbf{1. Sample} \\
    Sample target \(r^*_t\) from \\ Transition matrix
};
\node[box, below=of step1] (step2) {
    \textbf{2. Instruct} \\
    Build prompt: position context \\
    + role-specific instruction
};
\node[box, below=of step2] (step3) {
    \textbf{3. Generate} \\
    Model produces response
};
\node[box, right=of step3] (step4) {
    \textbf{4. Classify} \\
    Classify actual output \\
    \(\to\) observed \(\text{role}_t\)
};
\node[box, right=of step2] (step5) {
    \textbf{5. Feed back} \\
    Set \(\text{role}_{t-1} \leftarrow \text{role}_t\) \\
    Increment \(t\)
};
\draw[arrow] (step1) -- (step2);
\draw[arrow] (step2) -- (step3);
\draw[arrow] (step3) -- (step4);
\draw[arrow] (step4) -- (step5);
\draw[feedback] (step5.north) |- ++(0,0.9cm) -| (step1.east)
    node[pos=0.25, above, align=center,
         font=\sffamily\footnotesize\bfseries, text=black!80] {Observed state\\(not sampled target)};
\end{tikzpicture}%
}
\caption{Closed-loop Markov steering procedure. At each turn, the sampler draws a human-typical target role conditioned on the model's previous \emph{classified} role and conversation phase.}
\label{fig:steering-loop}
\end{wrapfigure}
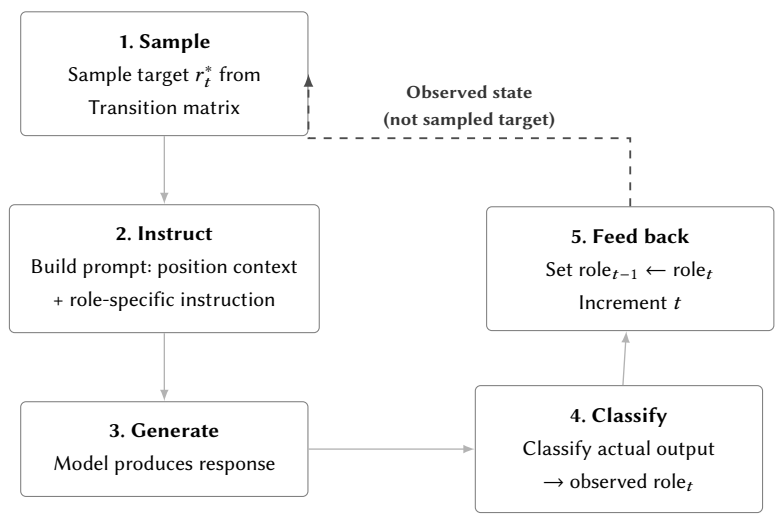

In \autoref{sec:results-rq2-transition} we noted the sequential dependence (\autoref{eq:rq3-lagged-model}  of \rolestates on prior ones. This dynamic effectively reflects a first-order Markov chain, which inspired us to consider the transition of \rolestates as Markovian process~ \cite{norris_markov_1997}. We leveraged this process to steer the \rolestates of AI. We use the \hh reference data to estimate the transition probabilities from \rolestate to another over early, middle, or late third phases. For instance, \RoleExploring $\rightarrow$ \RoleReflecting has an overall transition probability of $0.3$ in counselling context (\autoref{tab:trans-mi}). We used these transition distributions to nudge the AI to behave based on specific \rolestates at specific times in the conversation.  We term this method \textbf{Markov-informed Steering.} \autoref{fig:steering-loop} shows the Markov steering procedure.

To evaluate the effectiveness of our method, we compare three conditions, each representing a different level of structural guidance:
\begin{itemize}
  \item \textbf{Prompt}: A context-appropriate system prompt with no structural guidance. For example, in MI: \emph{``You are a motivational interviewing counsellor. The client is exploring a behaviour change. Respond empathetically and help them explore their ambivalence.''}
  \item \textbf{Unsteered Role}: The system prompt provides specific instructions on what support strategies to use and how to deliver it based on AROMA. For example: \emph{``You are a motivational interviewing counsellor. Use reflective listening extensively and vary your approach across turns: sometimes ask open questions, sometimes summarise, sometimes offer direction.''}
  \item \textbf{Markov}: This approach adds per-turn scaffolding to steer the AI supporter behavior through chain-sampled instructions. Before each turn, we sample a target role from the role transition matrix \autoref{tab:trans-mi}   conditioned on the previous classified role and conversation phase.
\end{itemize}

Using DeepSeek-v3 \cite{deepseek-ai_deepseek-v3_2025} we simulated 30 new \aiai conversations of 20 supporter turns per condition per care context (MI, Peer, and Counselling), yielding 270 conversations and 5,400 supporter turns. We compared the different supporter prompting strategies across these simulated conversations by measuring how different it is from \hh support from our corpora. To evaluate the effectiveness of using AROMA for support steering, we statistically evaluated two aspects of support behavior. First, we checked how AI supporter \rolestates align with that of human supporters using JS-Divergence. Second, we measured how AI supporters' generated language changed across the conditions by measuring different structural attributes and psycholinguistic attributes. 

\subsection{Results}

\begin{figure}[t]
\centering
\includegraphics[width=\linewidth]{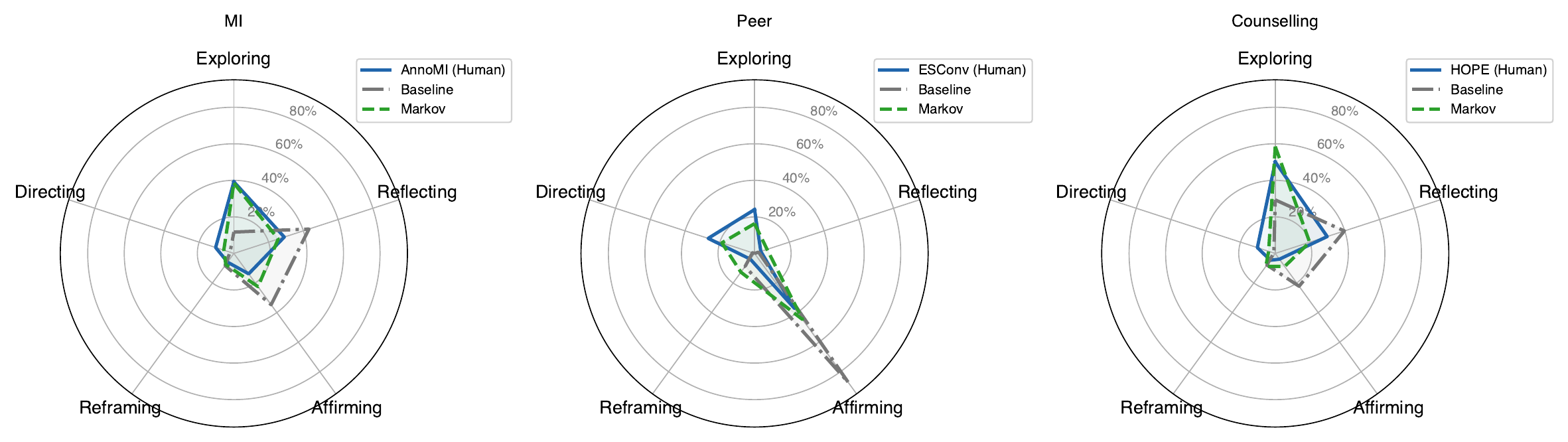}
\caption{Markov-informed steering (green) brings the AI distribution closest to the human reference (blue). Prompt (grey) and Unsteered Role (pink) remain further from the human target.}
\label{fig:radar-markov}
\Description{Three radar charts comparing human reference, Markov-steered, prompt-only, and unsteered AI role distributions for MI, Peer, and Counselling. Markov steering closely matches human profiles with JS divergence reductions of 70 to 85 percent.}
\end{figure}

\begin{figure}[t]
\centering
\includegraphics[width=\linewidth]{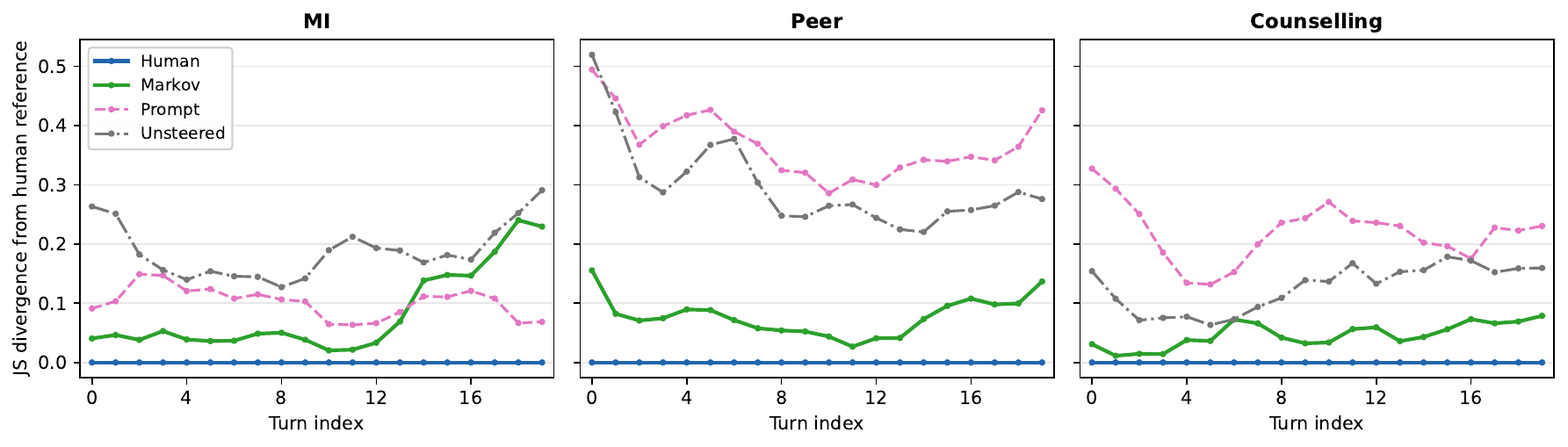}
\caption{JS divergence from the human reference at each turn position. Markov steering (green) maintains the lowest divergence throughout but drifts upward in late turns, indicating gradual loss of structural control. Prompt (grey, dash-dot) and Unsteered Role (pink, dashed) remain far from the human target at every position.}
\label{fig:js-over-turns}
\Description{Line chart showing Jensen-Shannon divergence from human reference at each turn position for three conditions across three domains. Markov steering maintains lowest divergence but drifts upward in late turns.}
\end{figure}

\subsubsection{Markov-steering leads to more aligned roles.}
Our comparative analysis found the Markov-informed steering significantly reduces the gap between AI and human role enactment. \autoref{fig:radar-markov} shows that steering moves the distribution of \rolestates towards the human reference more effectively than either the Prompt or Unsteered Role conditions. We observe this lower divergence across all contexts. In comparison to the Prompt condition, the  Markov-steered conversations exhibit 70--85\% lower JS divergence. By contrast, the Unsteered Role shows improvements over baseline prompts in peer and counseling conditions. However, the Markov-steered conversations reveal consistently lower divergence.  Furthermore, we inspected how enacted roles align across the conversation length. As shown in \autoref{fig:js-over-turns}, we found that although the divergence in the Markov-steered approach remains low for majority of the conversation, we notice a decay towards later stages. Together, Markov-steering shows improvements over the alternatives, but it also has challenges.


\subsubsection{AROMA's \rolestates help AI supporters match human linguistics.}
\begin{table*}[t]
  \centering
  \caption{This table presents 12 metrics across three categories: structure and style (verbosity, complexity, analytic thinking and linguistic accommodation, psycholinguistic indicators of speach (Linguistic Inquiry and Word Count), and BERT-scored empathy. Each value is a difference (AI minus human): positive values indicate the AI exceeds the human baseline; negative values indicate it falls below. \colorbox{green!15}{Green} and \colorbox{red!15}{red} highlighting indicate which AI condition is closest to or farthest from the human baseline per metric.}
  \label{tab:linguistic-delta-human}
   \Description{Table showing differences between AI and human linguistic profiles across 12 metrics for MI, Peer, and Counselling. Green cells indicate Markov steering moved closer to human baseline; red cells indicate divergence.}
 
  \scriptsize
  \begin{tabular}{@{} l r rrr | r rrr | r rrr @{}}
    \toprule
    & \multicolumn{4}{c|}{MI} & \multicolumn{4}{c|}{Peer} & \multicolumn{4}{c}{Counselling} \\
    \cmidrule(lr){2-5} \cmidrule(lr){6-9} \cmidrule(lr){10-13}
     & Hum. & Prompt & Unst. & Mark. & Hum. & Prompt & Unst. & Mark. & Hum. & Prompt & Unst. & Mark. \\
    \midrule
    \multicolumn{13}{@{}l}{\textit{Structure \& Style}} \\
    \quad Verbosity (words) & \textbf{24.4} & \cellcolor{red!15}$+$3.7 & \cellcolor{green!15}$-$0.8 & $-$2.3 & \textbf{15.4} & \cellcolor{red!15}$+$9.0 & $+$8.6 & \cellcolor{green!15}$+$6.1 & \textbf{27.1} & $-$2.9 & \cellcolor{green!15}$-$2.2 & \cellcolor{red!15}$-$6.8 \\
    \quad Complexity (word len.) & \textbf{4.07} & $+$0.30 & \cellcolor{red!15}$+$0.31 & \cellcolor{green!15}$+$0.25 & \textbf{3.96} & \cellcolor{red!15}$+$0.44 & \cellcolor{green!15}$+$0.30 & $+$0.36 & \textbf{4.22} & \cellcolor{red!15}$+$0.21 & $+$0.13 & \cellcolor{green!15}$+$0.04 \\
    \quad CDI (analytic) & \textbf{$-$13.9} & \cellcolor{red!15}$+$16.1 & \cellcolor{green!15}$+$15.0 & $+$15.4 & \textbf{$-$13.2} & $+$4.2 & \cellcolor{green!15}$+$2.6 & \cellcolor{red!15}$+$11.0 & \textbf{$-$14.3} & $+$15.5 & \cellcolor{red!15}$+$16.6 & \cellcolor{green!15}$+$10.6 \\
    \quad Ling.\ accommodation & \textbf{0.650} & \cellcolor{red!15}$+$0.140 & \cellcolor{green!15}$+$0.119 & $+$0.120 & \textbf{0.656} & \cellcolor{red!15}$+$0.171 & $+$0.163 & \cellcolor{green!15}$+$0.112 & \textbf{0.655} & $+$0.126 & \cellcolor{red!15}$+$0.127 & \cellcolor{green!15}$+$0.116 \\
    \midrule
    \multicolumn{13}{@{}l}{\textit{Pyscholingustics (TF-IDF of LIWC)}} \\
    \quad Affect (total) & \textbf{0.018} & $+$0.000 & $+$0.001 & $-$0.001 & \textbf{0.029} & \cellcolor{green!15}$-$0.004 & $-$0.005 & \cellcolor{red!15}$-$0.010 & \textbf{0.019} & $+$0.001 & $+$0.000 & $-$0.004 \\
    \quad Positive emotion & \textbf{0.031} & $-$0.014 & \cellcolor{green!15}$-$0.011 & \cellcolor{red!15}$-$0.014 & \textbf{0.044} & \cellcolor{red!15}$-$0.019 & \cellcolor{green!15}$-$0.013 & $-$0.019 & \textbf{0.023} & $-$0.007 & \cellcolor{green!15}$-$0.005 & \cellcolor{red!15}$-$0.009 \\
    \quad Negative emotion & \textbf{0.010} & \cellcolor{red!15}$+$0.023 & $+$0.020 & \cellcolor{green!15}$+$0.017 & \textbf{0.025} & \cellcolor{red!15}$+$0.017 & $+$0.006 & \cellcolor{green!15}$-$0.001 & \textbf{0.024} & \cellcolor{red!15}$+$0.017 & $+$0.010 & \cellcolor{green!15}$+$0.004 \\
    \quad Social (total) & \textbf{0.013} & $-$0.003 & $-$0.003 & $-$0.004 & \textbf{0.012} & $-$0.002 & $+$0.000 & $-$0.001 & \textbf{0.013} & $-$0.001 & $-$0.002 & $-$0.001 \\
    \quad Perceptual (total) & \textbf{0.015} & \cellcolor{red!15}$+$0.026 & \cellcolor{green!15}$+$0.019 & $+$0.020 & \textbf{0.015} & \cellcolor{red!15}$+$0.032 & \cellcolor{green!15}$+$0.015 & $+$0.022 & \textbf{0.015} & $+$0.028 & \cellcolor{red!15}$+$0.029 & \cellcolor{green!15}$+$0.020 \\
    \quad Feel & \textbf{0.011} & \cellcolor{red!15}$+$0.025 & \cellcolor{green!15}$+$0.017 & $+$0.023 & \textbf{0.011} & \cellcolor{red!15}$+$0.038 & \cellcolor{green!15}$+$0.017 & $+$0.021 & \textbf{0.013} & \cellcolor{red!15}$+$0.026 & $+$0.021 & \cellcolor{green!15}$+$0.016 \\
    \quad Biological (total) & \textbf{0.027} & $-$0.003 & $-$0.002 & $-$0.001 & \textbf{0.016} & \cellcolor{red!15}$+$0.017 & \cellcolor{green!15}$+$0.011 & $+$0.013 & \textbf{0.021} & $+$0.006 & $+$0.003 & $-$0.001 \\
    \quad Work & \textbf{0.021} & \cellcolor{green!15}$-$0.003 & $-$0.007 & \cellcolor{red!15}$-$0.009 & \textbf{0.029} & \cellcolor{red!15}$-$0.023 & $-$0.019 & \cellcolor{green!15}$-$0.016 & \textbf{0.031} & $-$0.020 & \cellcolor{green!15}$-$0.018 & \cellcolor{red!15}$-$0.020 \\
    \quad Assent & \textbf{0.048} & $-$0.047 & $-$0.045 & $-$0.045 & \textbf{0.022} & \cellcolor{green!15}$-$0.010 & $-$0.013 & \cellcolor{red!15}$-$0.016 & \textbf{0.034} & $-$0.033 & $-$0.033 & $-$0.033 \\
    \midrule
    \quad Empathy (BERT) & \textbf{0.823} & $+$0.047 & $+$0.046 & $+$0.049 & \textbf{0.752} & \cellcolor{red!15}$+$0.125 & $+$0.123 & \cellcolor{green!15}$+$0.118 & \textbf{0.886} & \cellcolor{red!15}$-$0.017 & $-$0.012 & \cellcolor{green!15}$-$0.004 \\
    \midrule
    \bottomrule
  \end{tabular}
  \begin{flushleft}\scriptsize
    Human $N$ (EN-only): MI\,=\,1{,}915, Peer\,=\,2{,}280, Counselling\,=\,1{,}900; synthetic $N = 600$ per condition.
    $\Delta$ = condition mean $-$ human mean. Sign shows direction of deviation.
  \end{flushleft}
\end{table*}

The differences in language use between the conditions is nuanced compared to our assessments of roles in the previous sections. Compared to the base prompt, the conditions leveraging AROMA's \rolestates (Unsteered and Markov-steering) tend to approximate human supporters' language better. However, the exact choice of approach can vary across contexts and depends on the metric we want to optimize for. 


\textbf{Unsteered roles tend to be more analytical; Markov-steered roles are more adaptive to seeker language}. In terms of structure, we found the Unsteered Roles to align with human supporters in terms of verbosity for MI and counselling contexts, and demonstrates similar analytical language  (measured by Categorical-dynamic index (CDI) \cite{pennebaker_when_2014}) for MI and peer contexts. Compared to humans, all AI supporters show higher CDI, which indicates a narrative style of speaking. Instead, research shows that lower CDI language, which is more descriptive and categorical tends to be more appropriate when responding to support seekers' personal challenges \cite{zhou_language_2023}. The Markov-steering showed better alignment in terms of the complexity of the words used (based on word-length) in MI and counselling as well as better linguistic accommodation in peer and counselling. The linguistic accommodation in Markov-steered AI support tends to show lower but closer scores to humans. This measure shows to what degree a supporter adapts their words to the seekers' speech. In general high adaptability is considered superior to template language \cite{chen_who_2026}. Having said that, over-adaptation can appear as mimicry which could appear insincere to support seekers \cite{cheng_sycophantic_2026}. 

\textbf{Unsteered roles match human's psycholingsuitics word-use; Markov-steered roles show more emphatic language use}. For the psycholinguistic aspects of speech, we calculated the importance of different support specific LIWC categories \cite{tausczik_psychological_2010, boyd_development_2022}   using \textit{TF-IDF} \cite{salton_term-weighting_1988}. We found Unsteered Roles to be closer to human supporters across multiple categories in the MI and peer context. However, Markov-steering showed better alignment across most categories in the counselling context. Specifically, Unsteered Roles were more successful in aligning in the use of positive emotion words (\autoref{tab:linguistic-delta-human}), while Markov-steering aligned in using negative emotion words (\autoref{tab:linguistic-delta-human}). Both sets of words have different functions in support, where positive words can be affirming and negative words can show acknowledgement (\autoref{tab:linguistic-delta-human}). Also note, we observed Unsteered Roles to show better alignment on perceptual (\autoref{tab:linguistic-delta-human}) and feeling words (\autoref{tab:linguistic-delta-human}) in MI and peer support, whereas Markov shows better alignment on these categories for counselling. These words can reflect how a supporter express that they have insight into the seeker's situation and are actively absorbing the information~\cite{pennebaker_when_2014}. More generally, for empathy (measured through BERT transformer \cite{reimers_sentence-bert_2019} ), we find the Markov-steering to show the best alignment. 

Therefore, depending on the contextual needs, the AI supporters can be tuned towards simulating human supporters with AROMA's \rolestates. On one hand, in contexts such as MI and peer support, Unsteered Roles can show alignment which paves the way for more efficient prompting. On the other hand, in contexts like counseling, Markov-steering's improved alignment can motivate the need to study role transitions and scaffold AI behavior at every turn.

\section{Discussion}

We presented a descriptive, validated taxonomy of eight Role States that uniquely distinguished supporter characteristics across motivational interviewing, counseling, and peer support, finding that specific roles were more prominently featured in supporters of a particular context (\textbf{RQ1}). 
We also found that AI supporters deviate both in the structure of their roles and in how they transition through roles compared to humans, and were often less dynamic (\textbf{RQ2}). Motivated by this deviation, we developed and evaluated a novel Markov-steering prompting approach that shifted AI's enacted role distributions more closely to humans. When assessed against human linguistics, prompting approaches that leveraged our taxonomy of Role States approximated human supporters' language better, but the best approach was nuanced and varied across contexts (\textbf{RQ3}). Reflecting together on our development of AROMA and on the implications of our findings, we offer two key points for discussion: (1) how Role States surface potential challenges of AI-driven support conversations, and (2) how different levels of abstraction shape our ability to analyze and steer AI support.


\subsection{Role States Surface Challenges in AI Support Systems}

Our findings suggest that examining support conversations through the lens of Role States could surface limitations of AI support systems. 
We found that AI roles differ from human roles, both in terms of role distribution and turn transition dynamics, even when comparing within a specific support context. AI roles tend to be more static, with less flexibility. This was notable as generative AI systems are often referenced by their flexibility to respond to a wide range of needs and adapt to user responses turn to turn. Yet, at the level of conversational roles, we observed the opposite. Thus, while an AI's language may appear adaptive at individual turns, its broader supportive role may remain comparatively static. Role States can offer a vocabulary for making these higher-level conversational patterns visible.

This rigidity matters because effective support may depend not only on performing an appropriate role, but also on recognizing when that role is no longer appropriate during the conversation and shifting accordingly. Human therapists are explicitly trained to attend to such dynamics, and even informal human peer supporters can draw on social and relational cues to adjust how they respond. Our findings suggest that AI systems may not make these shifts in the same ways. This distinction may become especially consequential when AI constitutes a person's primary or only accessible source of emotional support. Repeated exposure to a comparatively narrow set of Role States could shape expectations about what receiving support looks like, including expectations carried into future interactions with human supporters. This provides another reason to be cautious about framing AI mental health systems as therapists: doing so risks misrepresenting not simply the system's capabilities, but the nature and dynamics of the therapeutic relationship itself \cite{khawaja_your_2023}. Related work has similarly raised ethical risks surrounding the use of language models in automated mental healthcare \cite{grabb_risks_2024}. Sycophancy may compound these concerns: systems that default toward agreeable or socially desirable responses can decrease prosocial behavior and foster dependence \cite{cheng_sycophantic_2026}. For people who are particularly vulnerable or in need of human connection, role-level rigidity could risk orienting support increasingly inward toward continued AI interaction rather than outward toward other people.

Our analysis of both sides of the support interaction --- supporters and seekers --- surfaces a further concern: AI may shape not only how support is \emph{provided}, but also how support conversations \emph{unfold}. We found that AI seekers tended to shift conversations toward more action- and solution-oriented Role States. This pattern has implications for the growing use of AI--AI interactions to generate synthetic support conversations. If AI-generated seekers systematically steer conversations toward actionability, synthetic datasets produced through interactions between AI seekers and AI supporters may over-represent solution-oriented conversations relative to human ones. When such datasets are subsequently used to train or evaluate AI support systems, this bias could become self-reinforcing: AI seekers elicit more action-oriented support, those interactions enter training data, and future systems learn that effective support should itself be more action-oriented. Yet, it is common knowledge that people do not always want to hear solutions when engaging in support conversations. Effective support requires moving around a broader set of roles. Role States thus provides an opportunity for auditing the behavior of supportive AI systems, as well as the conversational assumptions that are embedded in the synthetic AI--AI datasets that are increasingly used to develop them.

\subsection{Choosing the Right Level of Abstraction for AI Support Conversations}

Our findings also raise a broader methodological question: \textit{At what level of abstraction, or what granularity, should we be looking to analyze and steer effective AI support?} This question becomes particularly relevant when situating our Role States and steering prompts alongside other approaches to analyzing and structuring AI support conversations.

Compared to our bottom-up characterization of \textit{enacted} roles, existing approaches often operate top-down, prescribing values, identities, or roles that an AI should follow. Yet, what a model is prescribed to be may differ from how those prescriptions manifest in conversation. Work on value alignment illustrates this distinction: ValueActionLens finds that what models \emph{say} about values can differ from how they \emph{act} in value-relevant situations \cite{shen_mind_2025}. Moreover, abstractions such as values can be inherently subjective and context-dependent: value alignment has been shown to vary across geographical contexts and models \cite{shen_mind_2025}, while ValueCompass demonstrates substantial variability across 56 human values \cite{shen_valuecompass_2024}. A bottom-up approach that instead is derived from observable characteristics of support conversations may offer a more concrete basis for analyzing what models actually do, rather than relying solely on whether their behavior conforms to broader prescribed abstractions.

Moving from values to prescribed roles reduces the level of abstraction, but still highlights important differences from our approach of enacted roles. Goel et al., for example, prescribe four roles drawn from social support theory---Inform, Coach, Relate, and Listen---and show that role assignment meaningfully shapes both perceived support quality and interactional risks \cite{goel_inform_2026}. Such top-down roles require strong theoretical grounding and allow designers to deliberately configure how an AI should provide support. However, specifying roles in advance may obscure distinctions or conversational dynamics that fall outside the chosen theoretical framework. Doing so also requires researchers to identify an appropriate theory and set of roles \emph{a priori}. Our finding that different support contexts exhibit distinct Role State distributions further highlights a limitation of prescribed roles: roles defined \emph{a priori} may not align with the dynamics of a particular support context, and may therefore overlook meaningful variation in how support is actually enacted. Conversely, a bottom-up approach could also be more context- or dataset- dependent, especially if derived from observed conversational patterns. The tradeoff may be centered around theoretical grounding for sensitivity to enacted, observable conversational behaviour.

Another option is to move to an even lower level of abstraction. Linguistic features can provide precise descriptions of how language is used, but may be difficult to translate into meaningful supportive actions or into prompts that can effectively steer AI support. Our work positions Role States as a potentially useful middle ground: high-level enough to represent meaningful supportive functions, yet low-level enough to observe and steer turn-by-turn. Rather than suggesting that roles are necessarily the ``right'' level of abstraction, our findings highlight a tradeoff among \textit{theoretical grounding, contextual sensitivity, and actionability}. One opportunity for future systems is therefore to analyze and steer support across multiple levels: values can guide \emph{what} good support should accomplish, roles can characterize \emph{what a supporter should do} at a particular moment, and linguistic features can shape \emph{how} that role is enacted.

\subsection{Limitations and Future Work}

Our work has several limitations that simultaneously open opportunities for future research. First, we do not claim that the Role States identified by AROMA are exhaustive. Our analysis was limited to the support contexts represented in publicly available datasets, which also predominantly captured text-based and single-session interactions. Other therapeutic approaches, support contexts, multi-session conversations, or even modalities such as voice, may surface Role States not captured here. Future work can apply and extend AROMA to these settings, both to assess the generalizability of our approach, the roles we identified, and to potentially surface new support roles.

Second, our evaluation of Role State steering focused on the linguistic properties of the steered conversations as an initial investigation, leaving an assessment of their effectiveness for human support seekers for future work. Producing conversations with desired Role State dynamics does not necessarily mean that those conversations provide better support. Future work should evaluate our Markov-based steering and prompting approaches with human support seekers and domain experts, including trained therapists or counselors, to examine whether steered Role States do indeed improve the perceived quality, appropriateness, and effectiveness of support.

Finally, although we used human support dynamics and roles as a benchmark for AI support in our analysis, human dynamics may not always be the optimal goal across contexts. Just as peers and therapists provide complementary forms of support, future work could examine how AI might similarly complement human support by exhibiting different Role States, rather than simply reproducing human patterns.

\section{Conclusion} 
We presented AROMA, a role-oriented framework for understanding the enacted social roles of AI supporters in mental well-being contexts. 
AROMA asks what roles an AI enacts, how those roles transition, and how they compare to human supporters across contexts. 
We found that support contexts have distinct Role State profiles, that AI supporters systematically underuse Reflecting and Affirming while switching roles less often than humans. We experimented with a Markov-informed steering mechanism that reduced divergence from human role profiles by 70–85\%. These findings position role alignment as a measurable design and evaluation target: not a replacement for user perception or clinical safety, but a structural complement that can reveal misalignment before it reshapes user expectations. Designers of AI support systems should consider not only what role a system is prompted to play, but what roles it enacts over time, how it transitions between them, and whether those transitions respect seeker autonomy. Future work should pursue three directions. First, steering should adapt dynamically to conversation-level variables such as seeker distress and topic sensitivity. Second, user-centric validation should establish whether structural alignment correlates with perceived empathy, therapeutic alliance, or well-being.  Thirdly, expand beyond the context of support by trying to ground other types of roles.
 
\bibliographystyle{ACM-Reference-Format}
\bibliography{references_edited , manul_reference}

@article{nytimes_tessa_2023,
  author    = {Hsu, Tiffany and Chung, Christine},
  title     = {A Wellness Chatbot Is Offline After Its 'Harmful' Focus on Weight Loss},
  journal   = {The New York Times},
  year      = {2023},
  month     = {June 8},
  url       = {https://www.nytimes.com/2023/06/08/us/ai-chatbot-tessa-eating-disorders-association.html},
  note      = {The artificial intelligence tool, named Tessa, was presented by the National Eating Disorders Association as a way to discover coping skills. But activists say it instead veered into problematic weight-loss advice.}
}

@inproceedings{reimers_sentence-bert_2019,
	title = {Sentence-{BERT}: {Sentence} {Embeddings} using {Siamese} {BERT}-{Networks}},
	shorttitle = {Sentence-{BERT}},
	url = {https://aclanthology.org/D19-1410/},
	doi = {10.18653/v1/D19-1410},
	urldate = {2026-09-11},
	booktitle = {Proceedings of the 2019 Conference on Empirical Methods in Natural Language Processing and the 9th International Joint Conference on Natural Language Processing (EMNLP-IJCNLP)},
	publisher = {Association for Computational Linguistics},
	address = {Hong Kong, China},
	author = {Reimers, Nils and Gurevych, Iryna},
	month = nov,
	year = {2019},
	pages = {3982--3992},
}

@article{oster_can_2026,
	title = {Can {Motivational} {Interviewing} be delivered using {Artificial} {Intelligence} chatbots? {Evaluating} the capability of {GPT}-4o},
	volume = {152},
	issn = {07383991},
	shorttitle = {Can {Motivational} {Interviewing} be delivered using {Artificial} {Intelligence} chatbots?},
	url = {https://linkinghub.elsevier.com/retrieve/pii/S0738399126003150},
	doi = {10.1016/j.pec.2026.109782},
	language = {en},
	urldate = {2026-09-11},
	journal = {Patient Education and Counseling},
	author = {Oster, Candice and Leibbrandt, Richard and Hocking, Judith and Muller, Amanda and Powers, David and Musolino, Renato and Lange, Belinda},
	month = nov,
	year = {2026},
	pages = {109782},
}

@article{jha_benchmarking_2026,
	title = {Benchmarking {Motivational} {Interviewing} {Competence} of {Large} {Language} {Models}},
	url = {https://karger.com/ear/article/doi/10.1159/000553455/951872/Benchmarking-Motivational-Interviewing-Competence},
	doi = {10.1159/000553455},
	urldate = {2026-09-11},
	journal = {European Addiction Research},
	publisher = {Karger},
	author = {Jha, Aishwariya and Shivaprakash, Prakrithi and Shukla, Lekhansh and Mukherjee, Animesh and Chand, Prabhat and Murthy, Pratima},
	month = jul,
	year = {2026},
	pages = {1--12},
}

@misc{zeng_who_2025,
	title = {Who is {In} {Charge}? {Dissecting} {Role} {Conflicts} in {Instruction} {Following}},
	copyright = {arXiv.org perpetual, non-exclusive license},
	shorttitle = {Who is {In} {Charge}?},
	url = {https://arxiv.org/abs/2510.01228},
	doi = {10.48550/ARXIV.2510.01228},
	urldate = {2026-09-11},
	publisher = {arXiv},
	author = {Zeng, Siqi},
	year = {2025},
}

@inproceedings{samuel_personagym_2025,
	address = {Suzhou, China},
	title = {{PersonaGym}: {Evaluating} {Persona} {Agents} and {LLMs}},
	isbn = {979-8-89176-335-7},
	url = {https://aclanthology.org/2025.findings-emnlp.368/},
	doi = {10.18653/v1/2025.findings-emnlp.368},
	booktitle = {Findings of the {Association} for {Computational} {Linguistics}: {EMNLP} 2025},
	publisher = {Association for Computational Linguistics},
	author = {Samuel, Vinay and Zou, Henry Peng and Zhou, Yue and Chaudhari, Shreyas and Kalyan, Ashwin and Rajpurohit, Tanmay and Deshpande, Ameet and Narasimhan, Karthik R and Murahari, Vishvak},
	editor = {Christodoulopoulos, Christos and Chakraborty, Tanmoy and Rose, Carolyn and Peng, Violet},
	month = nov,
	year = {2025},
	pages = {6999--7022},
}

@misc{xiao_when_2026,
	title = {When {Does} {Persona} {Prompting} {Actually} {Help}? {A} {Retrieval} and {Metric} {Analysis} of {Expert} {Role} {Injection} in {LLMs}},
	copyright = {arXiv.org perpetual, non-exclusive license},
	shorttitle = {When {Does} {Persona} {Prompting} {Actually} {Help}?},
	url = {https://arxiv.org/abs/2605.29420},
	doi = {10.48550/ARXIV.2605.29420},
	urldate = {2026-09-11},
	publisher = {arXiv},
	author = {Xiao, Shuai and Liu, Su and Zhou, Weikai and Wu, Jialun and He, Xinjie and Lin, Zhiyuan and Xie, Qiyang},
	year = {2026},
}

@article{greenberg_are_2006,
	title = {Are patient expectations still relevant for psychotherapy process and outcome?},
	volume = {26},
	issn = {02727358},
	url = {https://linkinghub.elsevier.com/retrieve/pii/S0272735805000255},
	doi = {10.1016/j.cpr.2005.03.002},
	language = {en},
	number = {6},
	urldate = {2026-09-11},
	journal = {Clinical Psychology Review},
	author = {Greenberg, Roger P. and Constantino, Michael J. and Bruce, Noah},
	month = oct,
	year = {2006},
	pages = {657--678},
}

@misc{pataranutaporn_my_2025,
	title = {"{My} {Boyfriend} is {AI}": {A} {Computational} {Analysis} of {Human}-{AI} {Companionship} in {Reddit}'s {AI} {Community}},
	shorttitle = {"{My} {Boyfriend} is {AI}"},
	url = {http://arxiv.org/abs/2509.11391},
	doi = {10.48550/arXiv.2509.11391},
	urldate = {2026-09-10},
	publisher = {arXiv},
	author = {Pataranutaporn, Pat and Karny, Sheer and Archiwaranguprok, Chayapatr and Albrecht, Constanze and Liu, Auren R. and Maes, Pattie},
	month = sep,
	year = {2025},
	note = {arXiv:2509.11391 [cs.HC]},
}

@article{tausczik_psychological_2010,
	title = {The {Psychological} {Meaning} of {Words}: {LIWC} and {Computerized} {Text} {Analysis} {Methods}},
	volume = {29},
	copyright = {https://journals.sagepub.com/page/policies/text-and-data-mining-license},
	issn = {0261-927X, 1552-6526},
	shorttitle = {The {Psychological} {Meaning} of {Words}},
	url = {https://journals.sagepub.com/doi/10.1177/0261927X09351676},
	doi = {10.1177/0261927X09351676},
	language = {en},
	number = {1},
	urldate = {2026-09-10},
	journal = {Journal of Language and Social Psychology},
	author = {Tausczik, Yla R. and Pennebaker, James W.},
	month = mar,
	year = {2010},
	pages = {24--54},
}

@inproceedings{shen_mind_2025,
	address = {Suzhou, China},
	title = {Mind the {Value}-{Action} {Gap}: {Do} {LLMs} {Act} in {Alignment} with {Their} {Values}?},
	shorttitle = {Mind the {Value}-{Action} {Gap}},
	url = {https://aclanthology.org/2025.emnlp-main.154},
	doi = {10.18653/v1/2025.emnlp-main.154},
	language = {en},
	urldate = {2026-09-10},
	booktitle = {Proceedings of the 2025 {Conference} on {Empirical} {Methods} in {Natural} {Language} {Processing}},
	publisher = {Association for Computational Linguistics},
	author = {Shen, Hua and Clark, Nicholas and Mitra, Tanu},
	year = {2025},
	pages = {3097--3118},
}

@incollection{turner_role-taking_1962,
	address = {Boston},
	title = {Role-{Taking}: {Process} {Versus} {Conformity}},
	booktitle = {Human {Behavior} and {Social} {Processes}},
	publisher = {Houghton Mifflin},
	author = {Turner, Ralph H.},
	editor = {Rose, Arnold M.},
	year = {1962},
	pages = {20--40},
}

@book{linton_study_1936,
	address = {New York},
	title = {The {Study} of {Man}: {An} {Introduction}},
	publisher = {Appleton-Century},
	author = {Linton, Ralph},
	year = {1936},
}

@inproceedings{plaza-del-arco_no_2025,
	address = {Suzhou, China},
	title = {No for {Some}, {Yes} for {Others}: {Persona} {Prompts} and {Other} {Sources} of {False} {Refusal} in {Language} {Models}},
	shorttitle = {No for {Some}, {Yes} for {Others}},
	url = {https://aclanthology.org/2025.winlp-main.39},
	doi = {10.18653/v1/2025.winlp-main.39},
	language = {en},
	urldate = {2026-09-10},
	booktitle = {Proceedings of the 9th {Widening} {NLP} {Workshop}},
	publisher = {Association for Computational Linguistics},
	author = {Plaza-del-Arco, Flor Miriam and Röttger, Paul and Scherrer, Nino and Borgonovo, Emanuele and Plischke, Elmar and Hovy, Dirk},
	year = {2025},
	pages = {268--282},
}

@inproceedings{tseng_two_2024,
	address = {Miami, Florida, USA},
	title = {Two {Tales} of {Persona} in {LLMs}: {A} {Survey} of {Role}-{Playing} and {Personalization}},
	shorttitle = {Two {Tales} of {Persona} in {LLMs}},
	url = {https://aclanthology.org/2024.findings-emnlp.969},
	doi = {10.18653/v1/2024.findings-emnlp.969},
	language = {en},
	urldate = {2026-09-10},
	booktitle = {Findings of the {Association} for {Computational} {Linguistics}: {EMNLP} 2024},
	publisher = {Association for Computational Linguistics},
	author = {Tseng, Yu-Min and Huang, Yu-Chao and Hsiao, Teng-Yun and Chen, Wei-Lin and Huang, Chao-Wei and Meng, Yu and Chen, Yun-Nung},
	year = {2024},
	pages = {16612--16631},
}

@article{salton_term-weighting_1988,
	title = {Term-weighting approaches in automatic text retrieval},
	volume = {24},
	copyright = {https://www.elsevier.com/tdm/userlicense/1.0/},
	issn = {03064573},
	url = {https://linkinghub.elsevier.com/retrieve/pii/0306457388900210},
	doi = {10.1016/0306-4573(88)90021-0},
	language = {en},
	number = {5},
	urldate = {2026-09-10},
	journal = {Information Processing \& Management},
	author = {Salton, Gerard and Buckley, Christopher},
	month = jan,
	year = {1988},
	pages = {513--523},
}

@misc{deepseek-ai_deepseek-v3_2025,
	title = {{DeepSeek}-{V3} {Technical} {Report}},
	url = {http://arxiv.org/abs/2412.19437},
	doi = {10.48550/arXiv.2412.19437},
	urldate = {2026-09-10},
	publisher = {arXiv},
	author = {DeepSeek-AI and Liu, Aixin and Feng, Bei and Xue, Bing and Wang, Bingxuan and Wu, Bochao and Lu, Chengda and Zhao, Chenggang and Deng, Chengqi and Zhang, Chenyu and Ruan, Chong and Dai, Damai and Guo, Daya and Yang, Dejian and Chen, Deli and Ji, Dongjie and Li, Erhang and Lin, Fangyun and Dai, Fucong and Luo, Fuli and Hao, Guangbo and Chen, Guanting and Li, Guowei and Zhang, H. and Bao, Han and Xu, Hanwei and Wang, Haocheng and Zhang, Haowei and Ding, Honghui and Xin, Huajian and Gao, Huazuo and Li, Hui and Qu, Hui and Cai, J. L. and Liang, Jian and Guo, Jianzhong and Ni, Jiaqi and Li, Jiashi and Wang, Jiawei and Chen, Jin and Chen, Jingchang and Yuan, Jingyang and Qiu, Junjie and Li, Junlong and Song, Junxiao and Dong, Kai and Hu, Kai and Gao, Kaige and Guan, Kang and Huang, Kexin and Yu, Kuai and Wang, Lean and Zhang, Lecong and Xu, Lei and Xia, Leyi and Zhao, Liang and Wang, Litong and Zhang, Liyue and Li, Meng and Wang, Miaojun and Zhang, Mingchuan and Zhang, Minghua and Tang, Minghui and Li, Mingming and Tian, Ning and Huang, Panpan and Wang, Peiyi and Zhang, Peng and Wang, Qiancheng and Zhu, Qihao and Chen, Qinyu and Du, Qiushi and Chen, R. J. and Jin, R. L. and Ge, Ruiqi and Zhang, Ruisong and Pan, Ruizhe and Wang, Runji and Xu, Runxin and Zhang, Ruoyu and Chen, Ruyi and Li, S. S. and Lu, Shanghao and Zhou, Shangyan and Chen, Shanhuang and Wu, Shaoqing and Ye, Shengfeng and Ye, Shengfeng and Ma, Shirong and Wang, Shiyu and Zhou, Shuang and Yu, Shuiping and Zhou, Shunfeng and Pan, Shuting and Wang, T. and Yun, Tao and Pei, Tian and Sun, Tianyu and Xiao, W. L. and Zeng, Wangding and Zhao, Wanjia and An, Wei and Liu, Wen and Liang, Wenfeng and Gao, Wenjun and Yu, Wenqin and Zhang, Wentao and Li, X. Q. and Jin, Xiangyue and Wang, Xianzu and Bi, Xiao and Liu, Xiaodong and Wang, Xiaohan and Shen, Xiaojin and Chen, Xiaokang and Zhang, Xiaokang and Chen, Xiaosha and Nie, Xiaotao and Sun, Xiaowen and Wang, Xiaoxiang and Cheng, Xin and Liu, Xin and Xie, Xin and Liu, Xingchao and Yu, Xingkai and Song, Xinnan and Shan, Xinxia and Zhou, Xinyi and Yang, Xinyu and Li, Xinyuan and Su, Xuecheng and Lin, Xuheng and Li, Y. K. and Wang, Y. Q. and Wei, Y. X. and Zhu, Y. X. and Zhang, Yang and Xu, Yanhong and Xu, Yanhong and Huang, Yanping and Li, Yao and Zhao, Yao and Sun, Yaofeng and Li, Yaohui and Wang, Yaohui and Yu, Yi and Zheng, Yi and Zhang, Yichao and Shi, Yifan and Xiong, Yiliang and He, Ying and Tang, Ying and Piao, Yishi and Wang, Yisong and Tan, Yixuan and Ma, Yiyang and Liu, Yiyuan and Guo, Yongqiang and Wu, Yu and Ou, Yuan and Zhu, Yuchen and Wang, Yuduan and Gong, Yue and Zou, Yuheng and He, Yujia and Zha, Yukun and Xiong, Yunfan and Ma, Yunxian and Yan, Yuting and Luo, Yuxiang and You, Yuxiang and Liu, Yuxuan and Zhou, Yuyang and Wu, Z. F. and Ren, Z. Z. and Ren, Zehui and Sha, Zhangli and Fu, Zhe and Xu, Zhean and Huang, Zhen and Zhang, Zhen and Xie, Zhenda and Zhang, Zhengyan and Hao, Zhewen and Gou, Zhibin and Ma, Zhicheng and Yan, Zhigang and Shao, Zhihong and Xu, Zhipeng and Wu, Zhiyu and Zhang, Zhongyu and Li, Zhuoshu and Gu, Zihui and Zhu, Zijia and Liu, Zijun and Li, Zilin and Xie, Ziwei and Song, Ziyang and Gao, Ziyi and Pan, Zizheng},
	month = feb,
	year = {2025},
	note = {arXiv:2412.19437 [cs.CL]},
}

@article{boyd_development_2022,
	title = {The {Development} and {Psychometric} {Properties} of {LIWC}-22},
	url = {https://www.researchgate.net/doi/10.13140/RG.2.2.23890.43205},
	doi = {10.13140/RG.2.2.23890.43205},
	language = {en},
	urldate = {2026-09-10},
	publisher = {Unpublished},
	author = {Boyd, Ryan L. and {Ashwini Ashokkumar} and Seraj, Sarah and Pennebaker, James W.},
	year = {2022},
}

@article{chen_who_2026,
	title = {Who {Accommodates} {Whom}? {Bidirectional} {Linguistic} {Accommodation} and {Progressive} {Interpersonal} {Convergence} in {Human}–{AI} {Conversations}},
	volume = {16},
	issn = {2076-328X},
	shorttitle = {Who {Accommodates} {Whom}?},
	url = {https://www.mdpi.com/2076-328X/16/5/720},
	doi = {10.3390/bs16050720},
	language = {en},
	number = {5},
	urldate = {2026-09-10},
	journal = {Behavioral Sciences},
	author = {Chen, Pengbo and Guan, Huining and Jeong, Eui Jun},
	month = may,
	year = {2026},
	pages = {720},
}

@book{norris_markov_1997,
	edition = {1},
	title = {Markov {Chains}},
	copyright = {https://www.cambridge.org/core/terms},
	isbn = {978-0-521-48181-6 978-0-521-63396-3 978-0-511-81063-3},
	url = {https://www.cambridge.org/core/product/identifier/9780511810633/type/book},
	doi = {10.1017/CBO9780511810633},
	urldate = {2026-09-10},
	publisher = {Cambridge University Press},
	author = {Norris, J. R.},
	month = feb,
	year = {1997},
}

@article{zhou_language_2023,
	title = {Language use in online support provision: the impact of others’ supportive messages, support-seeking strategy, and channel publicness},
	volume = {28},
	copyright = {https://creativecommons.org/licenses/by/4.0/},
	issn = {1083-6101},
	shorttitle = {Language use in online support provision},
	url = {https://academic.oup.com/jcmc/article/doi/10.1093/jcmc/zmac038/6982691},
	doi = {10.1093/jcmc/zmac038},
	language = {en},
	number = {2},
	urldate = {2026-09-10},
	journal = {Journal of Computer-Mediated Communication},
	author = {Zhou, Yanmengqian and High, Andrew C},
	editor = {Pearce, Katy},
	month = jan,
	year = {2023},
	pages = {zmac038},
}

@article{pennebaker_when_2014,
	title = {When {Small} {Words} {Foretell} {Academic} {Success}: {The} {Case} of {College} {Admissions} {Essays}},
	volume = {9},
	issn = {1932-6203},
	shorttitle = {When {Small} {Words} {Foretell} {Academic} {Success}},
	url = {https://dx.plos.org/10.1371/journal.pone.0115844},
	doi = {10.1371/journal.pone.0115844},
	language = {en},
	number = {12},
	urldate = {2026-09-10},
	journal = {PLoS ONE},
	author = {Pennebaker, James W. and Chung, Cindy K. and Frazee, Joey and Lavergne, Gary M. and Beaver, David I.},
	editor = {Gong, Qiyong},
	month = dec,
	year = {2014},
	pages = {e115844},
}

@article{lindgren_emerging_2025,
	title = {Emerging {Roles} and {Relationships} {Among} {Humans} and {Interactive} {AI} {Systems}},
	volume = {41},
	issn = {1044-7318, 1532-7590},
	url = {https://www.tandfonline.com/doi/full/10.1080/10447318.2024.2435693},
	doi = {10.1080/10447318.2024.2435693},
	language = {en},
	number = {17},
	urldate = {2026-09-10},
	journal = {International Journal of Human–Computer Interaction},
	author = {Lindgren, Helena},
	month = sep,
	year = {2025},
	pages = {10595--10617},
}

@misc{shin_roleconflictbench_2026,
	title = {{RoleConflictBench}: {A} {Benchmark} of {Role} {Conflict} {Scenarios} for {Evaluating} {LLMs}' {Contextual} {Sensitivity}},
	shorttitle = {{RoleConflictBench}},
	url = {http://arxiv.org/abs/2509.25897},
	doi = {10.48550/arXiv.2509.25897},
	urldate = {2026-09-10},
	publisher = {arXiv},
	author = {Shin, Jisu and Song, Hoyun and Oh, Juhyun and Ko, Changgeon and Kim, Eunsu and Jung, Chani and Oh, Alice},
	month = apr,
	year = {2026},
	note = {arXiv:2509.25897 [cs.CL]},
}

@misc{ding_contextecho_2026,
	title = {{ContextEcho}: {A} {Benchmark} for {Persona} {Drift} in {Long} {Agentic}-{Coding} {Sessions}},
	shorttitle = {{ContextEcho}},
	url = {http://arxiv.org/abs/2605.24279},
	doi = {10.48550/arXiv.2605.24279},
	urldate = {2026-09-10},
	publisher = {arXiv},
	author = {Ding, Xianzhong and Yu, Yangyang and Liu, Changwei and Zhao, Bill and Chen, Le and Chen, Tao},
	month = aug,
	year = {2026},
	note = {arXiv:2605.24279 [cs.CL]},
}

@inproceedings{song_role_2026,
	title = {Role {Steering} of {Language} {Models} for {Social} {Simulations}},
	url = {http://arxiv.org/abs/2608.00023},
	urldate = {2026-09-10},
	booktitle = {Proceedings of the {COLM} 2026 {Workshop} on {Social} {Simulations} with {LLMs}},
	author = {Song, Isaac and Parwani, Mohammed Rehan and Matlin, Glenn and Anand, Emile and Theerthala, Akhil and Chatterjee, Arjun and Zang, Anthony Wen-Ming and Kostylew, Maria and Shavit, Yonadav G. and Krier, Sebastien and Riedl, Mark},
	month = aug,
	year = {2026},
	note = {arXiv:2608.00023 [cs.CL]},
}

@inproceedings{lai_rolecdebenchmarking_2026,
	title = {{RoleCDE}:{Benchmarking} and {Mitigating} {Role}-{Alignment} {Trade}-offs in {Role}-{Playing} {Agents}},
	shorttitle = {{RoleCDE}},
	url = {https://aclanthology.org/2026.findings-acl.106/},
	doi = {10.18653/v1/2026.findings-acl.106},
	urldate = {2026-09-10},
	booktitle = {Findings of the Association for Computational Linguistics: {ACL} 2026},
	address = {San Diego, California, United States},
	publisher = {Association for Computational Linguistics},
	author = {Lai, Huayi and Song, Shichao and Niu, Simin and Wang, Hanyu and Yang, Jiawei and Wang, Zhouxing and Yin, Zhiqiang and Liang, Xun},
	month = jul,
	year = {2026},
}

@article{grabb_risks_2024,
	title = {Risks from {Language} {Models} for {Automated} {Mental} {Healthcare}: {Ethics} and {Structure} for {Implementation} ({Extended} {Abstract})},
	volume = {7},
	issn = {3065-8365},
	shorttitle = {Risks from {Language} {Models} for {Automated} {Mental} {Healthcare}},
	url = {https://ojs.aaai.org/index.php/AIES/article/view/31654},
	doi = {10.1609/aies.v7i1.31654},
	urldate = {2026-09-09},
	journal = {Proceedings of the AAAI/ACM Conference on AI, Ethics, and Society},
	author = {Grabb, Declan and Lamparth, Max and Vasan, Nina},
	month = oct,
	year = {2024},
	pages = {519--519},
}

@article{sharma_humanai_2023,
	title = {Human–{AI} collaboration enables more empathic conversations in text-based peer-to-peer mental health support},
	volume = {5},
	issn = {2522-5839},
	url = {https://www.nature.com/articles/s42256-022-00593-2},
	doi = {10.1038/s42256-022-00593-2},
	language = {en},
	number = {1},
	urldate = {2026-09-09},
	journal = {Nature Machine Intelligence},
	author = {Sharma, Ashish and Lin, Inna W. and Miner, Adam S. and Atkins, David C. and Althoff, Tim},
	month = jan,
	year = {2023},
	pages = {46--57},
}

@article{mccarthy_multitheoretical_2009,
	title = {The {Multitheoretical} {List} of {Therapeutic} {Interventions} ({MULTI}): {Initial} report},
	volume = {19},
	issn = {1050-3307, 1468-4381},
	shorttitle = {The {Multitheoretical} {List} of {Therapeutic} {Interventions} ({MULTI})},
	url = {http://www.tandfonline.com/doi/abs/10.1080/10503300802524343},
	doi = {10.1080/10503300802524343},
	language = {en},
	number = {1},
	urldate = {2026-09-08},
	journal = {Psychotherapy Research},
	author = {McCarthy, Kevin S. and Barber, Jacques P.},
	month = jan,
	year = {2009},
	pages = {96--113},
}

@inproceedings{nass_computers_1994,
	address = {Boston Massachusetts USA},
	title = {Computers are social actors},
	isbn = {978-0-89791-650-9},
	url = {https://dl.acm.org/doi/10.1145/191666.191703},
	doi = {10.1145/191666.191703},
	language = {en},
	urldate = {2026-09-07},
	booktitle = {Proceedings of the {SIGCHI} {Conference} on {Human} {Factors} in {Computing} {Systems}},
	publisher = {ACM},
	author = {Nass, Clifford and Steuer, Jonathan and Tauber, Ellen R.},
	month = apr,
	year = {1994},
	pages = {72--78},
}

@article{cheng_sycophantic_2026,
	title = {Sycophantic {AI} decreases prosocial intentions and promotes dependence},
	volume = {391},
	issn = {0036-8075, 1095-9203},
	url = {https://www.science.org/doi/10.1126/science.aec8352},
	doi = {10.1126/science.aec8352},
	language = {en},
	number = {6792},
	urldate = {2026-09-07},
	journal = {Science},
	author = {Cheng, Myra and Lee, Cinoo and Khadpe, Pranav and Yu, Sunny and Han, Dyllan and Jurafsky, Dan},
	month = mar,
	year = {2026},
	pages = {eaec8352},
}

@article{rautalinko_reflective_2013,
	title = {Reflective listening and open‐ended questions in counselling: {Preferences} moderated by social skills and cognitive ability},
	volume = {13},
	copyright = {http://onlinelibrary.wiley.com/termsAndConditions\#vor},
	issn = {1473-3145, 1746-1405},
	shorttitle = {Reflective listening and open‐ended questions in counselling},
	url = {https://onlinelibrary.wiley.com/doi/10.1080/14733145.2012.687387},
	doi = {10.1080/14733145.2012.687387},
	language = {en},
	number = {1},
	urldate = {2026-09-06},
	journal = {Counselling and Psychotherapy Research},
	author = {Rautalinko, Erik},
	month = mar,
	year = {2013},
	pages = {24--31},
}

@book{miller_motivational_2002,
	address = {New York},
	edition = {2. ed},
	title = {Motivational interviewing: preparing people for change},
	isbn = {978-1-57230-563-2},
	shorttitle = {Motivational interviewing},
	language = {eng},
	publisher = {Guilford Press},
	author = {Miller, William R. and Rollnick, Stephen},
	year = {2002},
}

@book{miller_motivational_1991,
	address = {New York},
	title = {Motivational interviewing: preparing people to change addictive behavior},
	isbn = {978-0-89862-566-0},
	shorttitle = {Motivational interviewing},
	language = {eng},
	publisher = {The Guilford Pr},
	editor = {Miller, William R. and Rollnick, Stephen},
	year = {1991},
}

@article{mead_peer_2001,
	title = {Peer support: {A} theoretical perspective.},
	volume = {25},
	issn = {1559-3126, 1095-158X},
	shorttitle = {Peer support},
	url = {https://doi.apa.org/doi/10.1037/h0095032},
	doi = {10.1037/h0095032},
	language = {en},
	number = {2},
	urldate = {2026-09-06},
	journal = {Psychiatric Rehabilitation Journal},
	author = {Mead, Shery and Hilton, David and Curtis, Laurie},
	year = {2001},
	pages = {134--141},
}

@inproceedings{gonnermann-muller_maintaining_2026,
	address = {Barcelona , Spain},
	title = {Maintaining {Stable} {Personas}? {Examining} {Temporal} {Stability} in {LLM}-{Based} {Human} {Simulation}},
	isbn = {979-8-4007-2281-3},
	shorttitle = {Maintaining {Stable} {Personas}?},
	url = {https://dl.acm.org/doi/10.1145/3772363.3799334},
	doi = {10.1145/3772363.3799334},
	language = {en},
	urldate = {2026-09-06},
	booktitle = {Proceedings of the {Extended} {Abstracts} of the 2026 {CHI} {Conference} on {Human} {Factors} in {Computing} {Systems}},
	publisher = {ACM},
	author = {Gonnermann-Müller, Jana and Haase, Jennifer and Leins, Nicolas and Kosch, Thomas and Pokutta, Sebastian},
	month = apr,
	year = {2026},
	pages = {1--6},
}

@misc{choi_examining_2025,
	title = {Examining {Identity} {Drift} in {Conversations} of {LLM} {Agents}},
	url = {http://arxiv.org/abs/2412.00804},
	doi = {10.48550/arXiv.2412.00804},
	urldate = {2026-09-06},
	publisher = {arXiv},
	author = {Choi, Junhyuk and Hong, Yeseon and Kim, Minju and Kim, Bugeun},
	month = feb,
	year = {2025},
	note = {arXiv:2412.00804 [cs.CY]},
}

@inproceedings{araujo_persistent_2026,
	title = {Persistent {Personas}? {Role}-{Playing}, {Instruction} {Following}, and {Safety} in {Extended} {Interactions}},
	shorttitle = {Persistent {Personas}?},
	url = {https://aclanthology.org/2026.eacl-long.246/},
	doi = {10.18653/v1/2026.eacl-long.246},
	urldate = {2026-09-06},
	booktitle = {Proceedings of the 19th Conference of the {E}uropean {C}hapter of the {A}ssociation for {C}omputational {L}inguistics (Volume 1: Long Papers)},
	address = {Rabat, Morocco},
	publisher = {Association for Computational Linguistics},
	author = {Luz de Araujo, Pedro Henrique and Hedderich, Michael A. and Modarressi, Ali and Schuetze, Hinrich and Roth, Benjamin},
	month = mar,
	year = {2026},
	pages = {5329--5359},
}

@misc{lu_assistant_2026,
	title = {The {Assistant} {Axis}: {Situating} and {Stabilizing} the {Default} {Persona} of {Language} {Models}},
	shorttitle = {The {Assistant} {Axis}},
	url = {http://arxiv.org/abs/2601.10387},
	doi = {10.48550/arXiv.2601.10387},
	urldate = {2026-09-06},
	publisher = {arXiv},
	author = {Lu, Christina and Gallagher, Jack and Michala, Jonathan and Fish, Kyle and Lindsey, Jack},
	month = jan,
	year = {2026},
	note = {arXiv:2601.10387 [cs.CL]},
}

@inproceedings{purohit_conditional_2026,
	address = {Barcelona Spain},
	title = {A {Conditional} {Companion}: {Lived} {Experiences} of {People} with {Mental} {Health} {Disorders} {Using} {LLMs}: {Conditional} {Companion}: {LLMs} \& {Mental} {Health}},
	isbn = {979-8-4007-2278-3},
	shorttitle = {A {Conditional} {Companion}},
	url = {https://dl.acm.org/doi/10.1145/3772318.3791763},
	doi = {10.1145/3772318.3791763},
	language = {en},
	urldate = {2026-09-06},
	booktitle = {Proceedings of the 2026 {CHI} {Conference} on {Human} {Factors} in {Computing} {Systems}},
	publisher = {ACM},
	author = {Purohit, Aditya Kumar and Heuer, Hendrik},
	month = apr,
	year = {2026},
	pages = {1--19},
}

@inproceedings{zhang_dark_2025,
	address = {Yokohama Japan},
	title = {The {Dark} {Side} of {AI} {Companionship}: {A} {Taxonomy} of {Harmful} {Algorithmic} {Behaviors} in {Human}-{AI} {Relationships}},
	isbn = {979-8-4007-1394-1},
	shorttitle = {The {Dark} {Side} of {AI} {Companionship}},
	url = {https://dl.acm.org/doi/10.1145/3706598.3713429},
	doi = {10.1145/3706598.3713429},
	language = {en},
	urldate = {2026-09-04},
	booktitle = {Proceedings of the 2025 {CHI} {Conference} on {Human} {Factors} in {Computing} {Systems}},
	publisher = {ACM},
	author = {Zhang, Renwen and Li, Han and Meng, Han and Zhan, Jinyuan and Gan, Hongyuan and Lee, Yi-Chieh},
	month = apr,
	year = {2025},
	pages = {1--17},
}

@inproceedings{jorke_gptcoach_2025,
	address = {Yokohama Japan},
	title = {{GPTCoach}: {Towards} {LLM}-{Based} {Physical} {Activity} {Coaching}},
	isbn = {979-8-4007-1394-1},
	shorttitle = {{GPTCoach}},
	url = {https://dl.acm.org/doi/10.1145/3706598.3713819},
	doi = {10.1145/3706598.3713819},
	language = {en},
	urldate = {2026-09-04},
	booktitle = {Proceedings of the 2025 {CHI} {Conference} on {Human} {Factors} in {Computing} {Systems}},
	publisher = {ACM},
	author = {Jörke, Matthew and Sapkota, Shardul and Warkenthien, Lyndsea and Vainio, Niklas and Schmiedmayer, Paul and Brunskill, Emma and Landay, James A.},
	month = apr,
	year = {2025},
	pages = {1--46},
}

@article{song_typing_2025,
	title = {The {Typing} {Cure}: {Experiences} with {Large} {Language} {Model} {Chatbots} for {Mental} {Health} {Support}},
	shorttitle = {The {Typing} {Cure}},
	url = {https://dl.acm.org/doi/10.1145/3757430},
	doi = {10.1145/3757430},
	urldate = {2026-09-04},
	journal = {Proceedings of the ACM on Human-Computer Interaction},
	volume = {9},
	number = {CSCW2},
	author = {Song, Inhwa and Pendse, Sachin R. and Kumar, Neha and Choudhury, Munmun De},
	month = oct,
	year = {2025},
	pages = {1--29},
	note = {Article CSCW249},
}

@misc{goel_inform_2026,
	title = {Inform, {Coach}, {Relate}, {Listen}: {Auditing} {LLM} {Caregiving} {Support} {Roles}},
	shorttitle = {Inform, {Coach}, {Relate}, {Listen}},
	url = {http://arxiv.org/abs/2605.29473},
	doi = {10.48550/arXiv.2605.29473},
	urldate = {2026-09-03},
	publisher = {arXiv},
	author = {Goel, Drishti and Goyal, Agam and Duddu, Veda and Pal, Olivia and Lee, Jeongah and Zhong, Qiuyue Joy and Rodriguez, Violeta J. and Brown, Daniel S. and Yoo, Dong Whi and Karkar, Ravi and Saha, Koustuv},
	month = jun,
	year = {2026},
	note = {arXiv:2605.29473 [cs.HC]},
}

@article{biddle_recent_1986,
	title = {Recent {Developments} in {Role} {Theory}},
	volume = {12},
	issn = {0360-0572, 1545-2115},
	url = {https://www.annualreviews.org/doi/10.1146/annurev.so.12.080186.000435},
	doi = {10.1146/annurev.so.12.080186.000435},
	language = {en},
	number = {1},
	urldate = {2026-08-31},
	journal = {Annual Review of Sociology},
	author = {Biddle, B. J.},
	month = aug,
	year = {1986},
	pages = {67--92},
}

@inproceedings{rashkin_towards_2019,
	address = {Florence, Italy},
	title = {Towards {Empathetic} {Open}-domain {Conversation} {Models}: {A} {New} {Benchmark} and {Dataset}},
	shorttitle = {Towards {Empathetic} {Open}-domain {Conversation} {Models}},
	url = {https://aclanthology.org/P19-1534},
	doi = {10.18653/v1/P19-1534},
	language = {en},
	urldate = {2026-08-29},
	booktitle = {Proceedings of the 57th {Annual} {Meeting} of the {Association} for {Computational} {Linguistics}},
	publisher = {Association for Computational Linguistics},
	author = {Rashkin, Hannah and Smith, Eric Michael and Li, Margaret and Boureau, Y-Lan},
	year = {2019},
	pages = {5370--5381},
}

@misc{yan_sharechat_2026,
	title = {{ShareChat}: {A} {Dataset} of {Chatbot} {Conversations} in the {Wild}},
	shorttitle = {{ShareChat}},
	url = {http://arxiv.org/abs/2512.17843},
	doi = {10.48550/arXiv.2512.17843},
	urldate = {2026-08-29},
	publisher = {arXiv},
	author = {Yan, Yueru and Nguyen, Tuc and Su, Bo and Lieffers, Melissa and Le, Thai},
	month = may,
	year = {2026},
	note = {arXiv:2512.17843 [cs.CL]},
}

@incollection{cutrona_type_1990,
	address = {New York},
	series = {Wiley series on personality processes},
	title = {Type of social support and specific stress: {Toward} a theory of optimal matching},
	isbn = {0-471-60624-3},
	booktitle = {Social {Support}: {An} {Interactional} {View}},
	publisher = {John Wiley \& Sons},
	author = {Cutrona, Carolyn E. and Russell, Daniel W.},
	editor = {Sarason, Barbara R. and Sarason, Irwin G. and Pierce, Gregory R.},
	year = {1990},
	pages = {319--366},
}

@article{suhSENSE7TaxonomyDataset2025,
	title = {{SENSE}-7: {Taxonomy} and {Dataset} for {Measuring} {User} {Perceptions} of {Empathy} in {Sustained} {Human}-{AI} {Conversations}},
	shorttitle = {{SENSE}-7},
	url = {https://www.microsoft.com/en-us/research/publication/sense-7-taxonomy-and-dataset-for-measuring-user-perceptions-of-empathy-in-sustained-human-ai-conversations/},
	urldate = {2026-08-17},
	journal = {IEEE Transactions on Affective Computing},
	author = {Suh, Jina and Le, Lindy and Shayegani, Erfan and Ramos, Gonzalo and Amores, Judith and Ong, Desmond C. and Czerwinski, Mary and Hernandez, Javier},
	month = feb,
	year = {2026},
	pages = {1--18},
}

@inproceedings{liuEmotionalSupportDialog2021,
	title = {Towards {Emotional} {Support} {Dialog} {Systems}},
	url = {https://aclanthology.org/2021.acl-long.269/},
	doi = {10.18653/v1/2021.acl-long.269},
	urldate = {2026-08-17},
	booktitle = {Proceedings of the 59th Annual Meeting of the Association for Computational Linguistics and the 11th International Joint Conference on Natural Language Processing (Volume 1: Long Papers)},
	address = {Online},
	publisher = {Association for Computational Linguistics},
	author = {Liu, Siyang and Zheng, Chujie and Demasi, Orianna and Sabour, Sahand and Li, Yu and Yu, Zhou and Jiang, Yong and Huang, Minlie},
	month = aug,
	year = {2021},
	pages = {3469--3483},
}

@misc{lee_cactus_2024,
	title = {Cactus: {Towards} {Psychological} {Counseling} {Conversations} using {Cognitive} {Behavioral} {Theory}},
	shorttitle = {Cactus},
	url = {http://arxiv.org/abs/2407.03103},
	doi = {10.48550/arXiv.2407.03103},
	urldate = {2026-08-17},
	publisher = {arXiv},
	author = {Lee, Suyeon and Kim, Sunghwan and Kim, Minju and Kang, Dongjin and Yang, Dongil and Kim, Harim and Kang, Minseok and Jung, Dayi and Kim, Min Hee and Lee, Seungbeen and Chung, Kyoung-Mee and Yu, Youngjae and Lee, Dongha and Yeo, Jinyoung},
	month = oct,
	year = {2024},
	note = {arXiv:2407.03103 [cs.CL]},
}

@inproceedings{wu_anno-mi_2022,
	address = {Singapore, Singapore},
	title = {Anno-{MI}: {A} {Dataset} of {Expert}-{Annotated} {Counselling} {Dialogues}},
	copyright = {https://doi.org/10.15223/policy-029},
	isbn = {978-1-6654-0540-9},
	shorttitle = {Anno-{MI}},
	url = {https://ieeexplore.ieee.org/document/9746035/},
	doi = {10.1109/ICASSP43922.2022.9746035},
	urldate = {2026-08-25},
	booktitle = {{ICASSP} 2022 - 2022 {IEEE} {International} {Conference} on {Acoustics}, {Speech} and {Signal} {Processing} ({ICASSP})},
	publisher = {IEEE},
	author = {Wu, Zixiu and Balloccu, Simone and Kumar, Vivek and Helaoui, Rim and Reiter, Ehud and Reforgiato Recupero, Diego and Riboni, Daniele},
	month = may,
	year = {2022},
	pages = {6177--6181},
}

@inproceedings{shen_valuecompass_2024,
	title = {{ValueCompass}: {A} {Framework} for {Measuring} {Contextual} {Value} {Alignment} {Between} {Human} and {LLMs}},
	shorttitle = {{ValueCompass}},
	url = {https://aclanthology.org/2025.winlp-main.15/},
	urldate = {2026-08-25},
	booktitle = {Proceedings of the 9th Widening {NLP} Workshop},
	address = {Suzhou, China},
	publisher = {Association for Computational Linguistics},
	author = {Shen, Hua and Knearem, Tiffany and Ghosh, Reshmi and Yang, Yu-Ju and Clark, Nicholas and Mitra, Tanushree and Huang, Yun},
	year = {2025},
	pages = {75--86},
}

@inproceedings{qiu_psydial_2025,
	address = {Vienna, Austria},
	title = {{PsyDial}: {A} {Large}-scale {Long}-term {Conversational} {Dataset} for {Mental} {Health} {Support}},
	shorttitle = {{PsyDial}},
	url = {https://aclanthology.org/2025.acl-long.1049},
	doi = {10.18653/v1/2025.acl-long.1049},
	language = {en},
	urldate = {2026-08-25},
	booktitle = {Proceedings of the 63rd {Annual} {Meeting} of the {Association} for {Computational} {Linguistics} ({Volume} 1: {Long} {Papers})},
	publisher = {Association for Computational Linguistics},
	author = {Qiu, Huachuan and Lan, Zhenzhong},
	year = {2025},
	pages = {21624--21655},
}

@inproceedings{qiu_interactive_2024,
	title = {Interactive {Agents}: {Simulating} {Counselor}-{Client} {Psychological} {Counseling} via {Role}-{Playing} {LLM}-to-{LLM} {Interactions}},
	shorttitle = {Interactive {Agents}},
	url = {https://aclanthology.org/2026.starsem-conference.29/},
	urldate = {2026-08-25},
	booktitle = {Proceedings of the 15th Joint Conference on Lexical and Computational Semantics (*{SEM} 2026)},
	address = {San Diego, California, United States},
	publisher = {Association for Computational Linguistics},
	author = {Qiu, Huachuan and Lan, Zhenzhong},
	month = jul,
	year = {2026},
}

@inproceedings{qi_kokorochat_2025,
	address = {Vienna, Austria},
	title = {{KokoroChat}: {A} {Japanese} {Psychological} {Counseling} {Dialogue} {Dataset} {Collected} via {Role}-{Playing} by {Trained} {Counselors}},
	shorttitle = {{KokoroChat}},
	url = {https://aclanthology.org/2025.acl-long.608},
	doi = {10.18653/v1/2025.acl-long.608},
	language = {en},
	urldate = {2026-08-25},
	booktitle = {Proceedings of the 63rd {Annual} {Meeting} of the {Association} for {Computational} {Linguistics} ({Volume} 1: {Long} {Papers})},
	publisher = {Association for Computational Linguistics},
	author = {Qi, Zhiyang and Kaneko, Takumasa and Takamizo, Keiko and Ukiyo, Mariko and Inaba, Michimasa},
	year = {2025},
	pages = {12424--12443},
}

@inproceedings{meyer_you_2024,
	address = {Sheffield United Kingdom},
	title = {"{You} tell me": {A} {Dataset} of {GPT}-4-{Based} {Behaviour} {Change} {Support} {Conversations}},
	isbn = {979-8-4007-0434-5},
	shorttitle = {"{You} tell me"},
	url = {https://dl.acm.org/doi/10.1145/3627508.3638330},
	doi = {10.1145/3627508.3638330},
	language = {en},
	urldate = {2026-08-25},
	booktitle = {Proceedings of the 2024 {ACM} {SIGIR} {Conference} on {Human} {Information} {Interaction} and {Retrieval}},
	publisher = {ACM},
	author = {Meyer, Selina and Elsweiler, David},
	month = mar,
	year = {2024},
	pages = {411--416},
}

@inproceedings{malhotra_speaker_2022,
	address = {Virtual Event AZ USA},
	title = {Speaker and {Time}-aware {Joint} {Contextual} {Learning} for {Dialogue}-act {Classification} in {Counselling} {Conversations}},
	isbn = {978-1-4503-9132-0},
	url = {https://dl.acm.org/doi/10.1145/3488560.3498509},
	doi = {10.1145/3488560.3498509},
	language = {en},
	urldate = {2026-08-25},
	booktitle = {Proceedings of the {Fifteenth} {ACM} {International} {Conference} on {Web} {Search} and {Data} {Mining}},
	publisher = {ACM},
	author = {Malhotra, Ganeshan and Waheed, Abdul and Srivastava, Aseem and Akhtar, Md Shad and Chakraborty, Tanmoy},
	month = feb,
	year = {2022},
	pages = {735--745},
}

@article{khawaja_your_2023,
	title = {Your robot therapist is not your therapist: understanding the role of {AI}-powered mental health chatbots},
	volume = {5},
	issn = {2673-253X},
	shorttitle = {Your robot therapist is not your therapist},
	url = {https://www.frontiersin.org/articles/10.3389/fdgth.2023.1278186/full},
	doi = {10.3389/fdgth.2023.1278186},
	urldate = {2026-08-25},
	journal = {Frontiers in Digital Health},
	author = {Khawaja, Zoha and Bélisle-Pipon, Jean-Christophe},
	month = nov,
	year = {2023},
	pages = {1278186},
}

@misc{chiu_computational_2024,
	title = {A {Computational} {Framework} for {Behavioral} {Assessment} of {LLM} {Therapists}},
	copyright = {arXiv.org perpetual, non-exclusive license},
	url = {https://arxiv.org/abs/2401.00820},
	doi = {10.48550/ARXIV.2401.00820},
	urldate = {2026-08-19},
	publisher = {arXiv},
	author = {Chiu, Yu Ying and Sharma, Ashish and Lin, Inna Wanyin and Althoff, Tim},
	year = {2024},
}

@article{chen_socialsim_2025,
	title = {{SocialSim}: {Towards} {Socialized} {Simulation} of {Emotional} {Support} {Conversation}},
	volume = {39},
	issn = {2374-3468, 2159-5399},
	shorttitle = {{SocialSim}},
	url = {https://ojs.aaai.org/index.php/AAAI/article/view/32116},
	doi = {10.1609/aaai.v39i2.32116},
	number = {2},
	urldate = {2026-08-19},
	journal = {Proceedings of the AAAI Conference on Artificial Intelligence},
	author = {Chen, Zhuang and Cao, Yaru and Bi, Guanqun and Wu, Jincenzi and Zhou, Jinfeng and Xiao, Xiyao and Chen, Si and Wang, Hongning and Huang, Minlie},
	month = apr,
	year = {2025},
	pages = {1274--1282},
}

@article{stiles_responsiveness_1998,
	title = {Responsiveness in psychotherapy.},
	volume = {5},
	copyright = {http://doi.wiley.com/10.1002/tdm\_license\_1.1},
	issn = {1468-2850, 0969-5893},
	url = {https://doi.apa.org/doi/10.1111/j.1468-2850.1998.tb00166.x},
	doi = {10.1111/j.1468-2850.1998.tb00166.x},
	language = {en},
	number = {4},
	urldate = {2026-07-29},
	journal = {Clinical Psychology: Science and Practice},
	author = {Stiles, William B. and Honos-Webb, Lara and Surko, Michael},
	year = {1998},
	pages = {439--458},
}

@book{hill_helping_2010,
	address = {Washington, DC},
	edition = {3. ed., 3. print},
	title = {Helping skills: facilitating exploration, insight, and action},
	isbn = {978-1-4338-0451-9},
	shorttitle = {Helping skills},
	language = {eng},
	publisher = {American Psychological Assoc},
	author = {Hill, Clara E.},
	year = {2010},
}

@inproceedings{das_swain_ai_2025,
	address = {Yokohama Japan},
	title = {{AI} on {My} {Shoulder}: {Supporting} {Emotional} {Labor} in {Front}-{Office} {Roles} with an {LLM}-based {Empathetic} {Coworker}},
	isbn = {979-8-4007-1394-1},
	shorttitle = {{AI} on {My} {Shoulder}},
	url = {https://dl.acm.org/doi/10.1145/3706598.3713705},
	doi = {10.1145/3706598.3713705},
	language = {en},
	urldate = {2026-05-29},
	booktitle = {Proceedings of the 2025 {CHI} {Conference} on {Human} {Factors} in {Computing} {Systems}},
	publisher = {ACM},
	author = {Das Swain, Vedant and Zhong, Qiuyue "Joy" and Parekh, Jash Rajesh and Jeon, Yechan and Zimmermann, Roy and Czerwinski, Mary P and Suh, Jina and Mishra, Varun and Saha, Koustuv and Hernandez, Javier},
	month = apr,
	year = {2025},
	pages = {1--29},
}

@article{moyers_motivational_2016,
	title = {The {Motivational} {Interviewing} {Treatment} {Integrity} {Code} ({MITI} 4): {Rationale}, {Preliminary} {Reliability} and {Validity}},
	volume = {65},
	issn = {07405472},
	shorttitle = {The {Motivational} {Interviewing} {Treatment} {Integrity} {Code} ({MITI} 4)},
	url = {https://linkinghub.elsevier.com/retrieve/pii/S0740547216000143},
	doi = {10.1016/j.jsat.2016.01.001},
	language = {en},
	urldate = {2026-05-20},
	journal = {Journal of Substance Abuse Treatment},
	author = {Moyers, Theresa B. and Rowell, Lauren N. and Manuel, Jennifer K. and Ernst, Denise and Houck, Jon M.},
	month = jun,
	year = {2016},
	pages = {36--42},
}

@inproceedings{shen_hidden_2026,
	title = {The {Hidden} {Puppet} {Master}: {Predicting} {Human} {Belief} {Change} in {Manipulative} {LLM} {Dialogues}},
	shorttitle = {The {Hidden} {Puppet} {Master}},
	url = {http://arxiv.org/abs/2603.20907},
	urldate = {2026-03-27},
	booktitle = {Conference on Language Modeling ({COLM})},
	author = {Shen, Jocelyn and Luvsanchultem, Amina and Kim, Jessica and Smith, Kynnedy and Danry, Valdemar and Rogers, Kantwon and Alghowinem, Sharifa and Park, Hae Won and Sap, Maarten and Breazeal, Cynthia},
	month = mar,
	year = {2026},
	note = {arXiv:2603.20907 [cs]},
}

@misc{manoli_shes_2025,
	title = {"{She}'s {Like} a {Person} but {Better}": {Characterizing} {Companion}-{Assistant} {Dynamics} in {Human}-{AI} {Relationships}},
	shorttitle = {"{She}'s {Like} a {Person} but {Better}"},
	url = {http://arxiv.org/abs/2510.15905},
	doi = {10.48550/arXiv.2510.15905},
	urldate = {2025-12-25},
	publisher = {arXiv},
	author = {Manoli, Aikaterina and Pauketat, Janet V. T. and Ladak, Ali and Noh, Hayoun and Hwang, Angel Hsing-Chi and Anthis, Jacy Reese},
	month = dec,
	year = {2025},
	note = {arXiv:2510.15905 [cs]},
}


\appendix


\section{Codebook Quick Reference}
\label{app:codebook}

This appendix condenses the AROMA codebook into a quick reference for the D1, D2, and D3 coding dimensions. The full codebook, with worked examples and boundary cases, is included in the replication package.

Each supporter turn is coded within its three-turn triad: seeker turn ($S_N$), supporter turn ($P_{N+1}$), seeker response ($S_{N+2}$). All three dimensions are coded independently.

\begin{table}[h]
  \centering
  \caption{AROMA coding dimensions.}
  \label{tab:dimensions}
  \begin{tabular}{@{} l l l l l @{}}
    \toprule
    Code & Dimension & Type & Values & Source \\
    \midrule
    D1 & Spans & Multi-label & Eight move labels & Hill (2010) \\
    D2 & Orientation & Five-class & Ask / RB / Ack / Adv / Other & Hill (2010), collapsed \\
    D3 & Support type & Binary & Nurturant / Action-Facilitating & Cutrona \& Russell (1990) \\
    \bottomrule
  \end{tabular}
\end{table}

\para{D1 span labels.} Tag every functionally distinct segment of the supporter's turn: \textit{Asking} (questions that open space), \textit{Restatement} (rephrases content), \textit{Reflection of Feelings} (mirrors emotion), \textit{Interpreting} (offers a new frame), \textit{Challenging} (points out discrepancies), \textit{Approval \& Reassurance} (validates or normalises), \textit{Advising} (suggests actions or provides information), and \textit{Personal Experience} (shares the supporter's own experience). D1 grounds D2.

\para{D2 orientation decision pipeline.}

{\small
\begin{verbatim}
Step 1.  Not a helping skill?             -> OTHER. Stop.
Step 2.  Tag D1 spans.
Step 3.  Trace details from S_N.
Step 4.  Contains a question?
           Exploratory (past/perception)  -> ASKING
           Directive (future action)      -> ADVISING
           Confirmation tag               -> continue
Step 5.  Restatement + question?
           S_N+2 answers question         -> ASKING
           S_N+2 confirms mirrored        -> REPEATING BACK
Step 6.  Verbatim echo or direct
         feeling reflection?              -> REPEATING BACK
Step 7.  2+ traced details?
           Substitution passes (backdrop) -> ACKNOWLEDGEMENT
           Substitution fails (load-bearing) -> REPEATING BACK
Step 8.  Affirmation of strength/
         progress/coping?                 -> ACKNOWLEDGEMENT
Step 9.  Content delivery?
           Remove advice -> nothing left  -> ADVISING
           Validation carries the turn    -> ACKNOWLEDGEMENT
           Extended advice (lists, plans) -> ADVISING
           Encouragement imperative       -> ACKNOWLEDGEMENT
Step 10. D1<->D2 consistency check.
Step 11. S_N+2 tie-breaker if needed.
\end{verbatim}
}

\para{Key distinctions.} \Asking opens the seeker's material further. \RB gives it back, processed: mirrors, rephrases, names. \Ack normalises, validates, or affirms without mirroring specifics. \Adv sets the seeker's material aside and delivers something new.

\para{Substitution test (RB vs.\ Ack).} Replace the $S_N$ details with unrelated details. If the turn still serves the same function, the details are backdrop $\to$ \Ack. If the turn collapses, the details are load-bearing $\to$ \RB.

\para{D3 support type decision pipeline.} Does the supporter introduce a new strategy, resource, or fact the seeker could use to act upon the stressor? If no, \Nur. If yes, remove the fact. If the turn still works relationally (fact is backdrop), \Nur. If the turn collapses (fact IS the turn), \AF.

\begin{table}[h]
  \centering
  \caption{Support subtypes mapped to D3 binary.}
  \label{tab:d3-subtypes}
  \begin{tabular}{@{} l l l @{}}
    \toprule
    Subtype & D3 & Attends to \\
    \midrule
    Emotional & Nurturant & Feelings, comfort \\
    Esteem    & Nurturant & Worth, competence \\
    Network   & Nurturant & Belonging, solidarity \\
    Informational & Action-Facilitating & Advice, guidance, facts \\
    Tangible  & Action-Facilitating & Concrete resources \\
    \bottomrule
  \end{tabular}
\end{table}


\section{Classifier Validation}
\label{app:classifier}

\subsection{Inter-Rater Reliability}

Reliability was assessed on 100 segments stratified across 17 domains, coded independently by two human raters and the DeepSeek v7.5.2 classifier.

\begin{table}[h]
  \centering
  \caption{D2 orientation: Cohen's $\kappa$ across rater pairs.}
  \label{tab:irr-d2}
  \begin{tabular}{@{} l r r r @{}}
    \toprule
    Pair & $\kappa$ & $P_o$ & $N$ \\
    \midrule
    Human--Human      & 0.713 & 79.0\% & 100 \\
    Human$_1$--LLM    & 0.686 & 77.0\% & 100 \\
    Human$_2$--LLM    & 0.587 & 69.0\% & 100 \\
    Consensus--LLM    & 0.734 & 81.0\% &  79 \\
    \bottomrule
  \end{tabular}
\end{table}

\begin{table}[h]
  \centering
  \caption{D3 support type: Gwet's AC1 across rater pairs. Cohen's $\kappa$ is inappropriate for D3 due to extreme prevalence skew (${\sim}$83\% nurturant); Gwet's AC1 corrects for this with a prevalence-robust chance model.}
  \label{tab:irr-d3}
  \begin{tabular}{@{} l r r r r @{}}
    \toprule
    Pair & AC1 & $P_o$ & $N$ & $\kappa$ \\
    \midrule
    Human--Human     & 0.740 & 81.2\% &  96 & 0.333 \\
    Human$_1$--LLM   & 0.774 & 83.0\% & 100 & 0.318 \\
    Human$_2$--LLM   & 0.773 & 84.4\% &  96 & 0.499 \\
    Consensus--LLM   & 0.889 & 91.0\% &  78 & 0.540 \\
    \bottomrule
  \end{tabular}
\end{table}


\section{D2 Classifier Prompt}
\label{app:d2-classifier-prompt}

The following system prompt was used for the DeepSeek-V3 classifier that annotated the full corpus. The classifier receives a three-turn triad (seeker $\to$ supporter $\to$ seeker response) and returns structured JSON with intermediate evidence fields.

\begin{small}
\begin{verbatim}
You are an expert in care conversation analysis,
trained in Hill's (2009) helping process model and
Cutrona & Russell's (1990) optimal matching theory.

## Task

Read a three-turn care interaction
(seeker S_N -> supporter P_N+1 -> seeker response S_N+2).
Code the SUPPORTER's turn using the three-phase pipeline
below. Output structured JSON with intermediate evidence
fields.

## Five orientations (D2)

- Asking: Questions/prompts that open a direction the
  seeker did not produce.
- Repeating back: Gives back what the seeker said,
  describing, rephrasing, naming their experience.
  The traced content IS the work.
- Acknowledgement: Presence without processing specifics,
  validation, normalising, esteem-building, solidarity.
- Advising: Delivers content the seeker didn't produce,
  advice, information, direction, plans, or
  change-pushing reframes.
- Other: Not a helping skill: institutional authority,
  session management, admin speech, AI error handling.

## PHASE 1: OBSERVE (tag spans, trace details)

1.1 Scope check. Non-helping-skill -> OTHER.
1.2 Tag D1 spans (asking, restatement, reflection,
    interpreting, challenging, approval, advising,
    personal_experience).
1.3 Trace details. List specific circumstantial details
    from S_N that the supporter references. Generic
    empathy traces nothing -> empty list.

## PHASE 2: ANALYSE (classify questions, verb intent)

2.1 Question Decision Matrix:
  Past/History -> ASKING
  Perception -> ASKING
  Future/Directive -> ADVISING
  Confirmation tag -> IGNORE (part of restatement)
  Admin/Factual -> OTHER

2.1b Mixed turns: Restates 2+ details AND asks a new
  question on same topic: seeker answers question ->
  ASKING; seeker confirms restatement -> REPEATING BACK.

2.2 Verb-Intent Classification:
  Evaluative (validate, encourage) -> BACKDROP
  Descriptive (restate, reflect) -> LOAD-BEARING
  Prescriptive (suggest, advise) -> delivering new content
  Sharing (personal experience) -> presence, not processing

## PHASE 3: DECIDE (first matching rule, stop)

1. Non-helping-skill -> OTHER
2. Asking + past/perception question -> ASKING
2b. Asking + past/perception + evaluative preamble
    -> ASKING
3. Asking + future/directive question -> ADVISING
4. Advising/challenging dominant -> ADVISING
5. Verbatim echo or direct mirror of named emotion
   -> REPEATING BACK
6. Only restatement/reflection, traced details present,
   descriptive verb -> REPEATING BACK
7. Approval or evaluative verb -> ACKNOWLEDGEMENT
8. Traced details empty + validation -> ACKNOWLEDGEMENT
9. Strip prescriptive: nothing remains -> ADVISING;
   validation carries turn -> ACKNOWLEDGEMENT
10. Default -> ACKNOWLEDGEMENT

## Output format

{"traced_details": [...],
 "question_type": "past|future|perception|...",
 "verb_intent": "evaluative|descriptive|...",
 "decision_rule": "<Phase 3 rule>",
 "description": "<1-2 sentences>",
 "D1_labels": [...],
 "D2_orientation": "asking|repeating_back|..."}
\end{verbatim}
\end{small}

The prompt was accompanied by 15 few-shot examples covering boundary cases (validation preamble + question, cognitive reframe vs.\ emotional encouragement, pure leave-taking, and mixed restatement-question turns).


\section{D3 Classifier Prompt}
\label{app:d3-classifier-prompt}

The following system prompt was used for the independent D3 support-type classifier (prompt version 6, DeepSeek-V3). The classifier receives only the supporter's turn text and returns structured JSON. It does not see D2 labels.

\begin{small}
\begin{verbatim}
You are a social support researcher classifying
supporter turns using Cutrona & Russell's (1990)
optimal matching theory.

## Theoretical foundation

- Nurturant support promotes emotion-focused coping:
  feeling heard, validated, less alone, more competent.
- Action-facilitating support promotes problem-focused
  coping: giving the person something to act on.

## The Five Support Types (SSBC)

Nurturant (attends to WHO THE PERSON IS):
1. Emotional: caring, empathy, comfort.
2. Esteem: validates worth/competence without
   prescribing action.
3. Network: solidarity, shared experience.

Action-facilitating (attends to WHAT THEY COULD DO):
4. Informational: advice, guidance, facts, referrals.
5. Tangible: concrete resources or direct help.

## Decision Procedure

Step 1 -- Read the supporter's turn only.

Step 2 -- Functional novelty test: does the supporter
  introduce a strategy, resource, or factual claim the
  seeker could use to act upon the stressor?
  YES -> proceed to Step 2b.
  NO  -> likely nurturant. Proceed to Step 3.

Step 2b -- Relational function check: when facts ARE
  present, what is the primary communicative function?
  Test: remove the factual content. Does the turn still
  accomplish its relational purpose?
  YES (fact is backdrop)  -> nurturant
  NO  (fact IS the turn)  -> action_facilitating

Crucial nuance: active listening is emotional support.
  "Have you noticed any not-so-good things about the
   smoking?" -> explores THEIR experience -> nurturant
  "Have you tried therapy?" -> introduces a SPECIFIC
   NEW option -> action_facilitating

Step 3 -- Identify primary support type (1-5).

Step 4 -- Derive higher-order category:
  Emotional/Esteem/Network -> nurturant
  Informational/Tangible   -> action_facilitating

## Output format

{"reasoning": "<1-2 sentences>",
 "novelty_test": "<yes/no and what>",
 "support_type": "emotional|esteem|network|
                  informational|tangible",
 "D3_support_type": "nurturant|action_facilitating"}
\end{verbatim}
\end{small}


\section{Multinomial Model Tables}
\label{app:multinomial}

Full coefficient tables for all multinomial logistic regressions reported in the paper. Reference category: \AskNur. Data: 22{,}785 segments, 16{,}760 transitions, 6{,}025 conversations, 12 domains.

\begin{table}[h]
  \centering
  \caption{Model summary. All models fit on the stratified analytical sample.}
  \label{tab:model-summary}
  \begin{tabular}{@{} l r r r @{}}
    \toprule
    Model & $R^2$ & LR $\chi^2$ & df \\
    \midrule
    RQ1: role $\sim$ context            & 0.071 & 2{,}387.7 & 14 \\
    RQ2: role $\sim$ seeker + supporter + context + length + position & 0.169 & --- & --- \\
    Lagged (base)                       & 0.174 & --- & --- \\
    Lagged (+prev\_role)                & 0.201 & --- & --- \\
    $\Delta$ (lagged)                   & 0.027 & 1{,}401.8 & 49 \\
    \bottomrule
  \end{tabular}
  \begin{flushleft}\small Omnibus context test: $\Delta$LR $\chi^2 = 4{,}046.5$, $\Delta$df $= 14$, $p < 10^{-4}$.\end{flushleft}
\end{table}

\begin{table*}[t]
  \centering
  \caption{RQ2 coefficients: role $\sim$ seeker\_type + supporter\_type + context + length + position. Reference levels: \AskNur (outcome), Human (seeker/supporter), Counselling (context). Cluster-robust standard errors ($B = 500$) in parentheses.}
  \label{tab:rq2-coefficients}
  \begin{tabular}{@{} l rrrrrrr @{}}
    \toprule
    Role & Intercept & Seeker$_\text{AI}$ & Supporter$_\text{AI}$ & MI & Peer & Length & Position \\
    \midrule
    \AckAF  & $-8.25$ {\scriptsize(0.47)} & $0.84$ {\scriptsize(0.40)} & $-2.98$ {\scriptsize(0.47)} & $0.87$ {\scriptsize(0.91)} & $3.44$ {\scriptsize(0.38)} & $0.011$ {\scriptsize(0.002)} & $3.25$ {\scriptsize(0.39)} \\
    \AckNur & $-3.32$ {\scriptsize(0.11)} & $0.36$ {\scriptsize(0.10)} & $-1.11$ {\scriptsize(0.09)} & $0.85$ {\scriptsize(0.12)} & $2.24$ {\scriptsize(0.07)} & $0.005$ {\scriptsize(0.001)} & $2.44$ {\scriptsize(0.09)} \\
    \AdvAF  & $-4.34$ {\scriptsize(0.12)} & $0.54$ {\scriptsize(0.11)} & $-0.61$ {\scriptsize(0.10)} & $1.08$ {\scriptsize(0.16)} & $1.91$ {\scriptsize(0.08)} & $0.015$ {\scriptsize(0.001)} & $2.67$ {\scriptsize(0.09)} \\
    \AdvNur & $-4.50$ {\scriptsize(0.14)} & $-0.01$ {\scriptsize(0.11)} & $-0.45$ {\scriptsize(0.10)} & $1.26$ {\scriptsize(0.12)} & $0.77$ {\scriptsize(0.10)} & $0.003$ {\scriptsize(0.001)} & $3.08$ {\scriptsize(0.13)} \\
    \AskAF  & $-5.10$ {\scriptsize(0.19)} & $0.41$ {\scriptsize(0.15)} & $0.20$ {\scriptsize(0.14)} & $0.85$ {\scriptsize(0.19)} & $0.26$ {\scriptsize(0.17)} & $0.011$ {\scriptsize(0.001)} & $1.35$ {\scriptsize(0.20)} \\
    \RBAF   & $-7.51$ {\scriptsize(1.10)} & $-0.41$ {\scriptsize(3.27)} & $-1.98$ {\scriptsize(2.62)} & $0.98$ {\scriptsize(1.98)} & $1.42$ {\scriptsize(1.06)} & $0.004$ {\scriptsize(0.009)} & $1.56$ {\scriptsize(0.59)} \\
    \RBNur  & $-1.49$ {\scriptsize(0.12)} & $0.22$ {\scriptsize(0.09)} & $-2.37$ {\scriptsize(0.08)} & $1.24$ {\scriptsize(0.11)} & $-0.27$ {\scriptsize(0.10)} & $0.005$ {\scriptsize(0.002)} & $0.59$ {\scriptsize(0.09)} \\
    \bottomrule
  \end{tabular}
\end{table*}

\para{Variance decomposition.} Context alone explains $R^2 = 0.067$ (given partner), while partner explains $R^2 = 0.009$ (given context)---context is 7.3$\times$ more predictive. Prior role is the single largest predictor ($\Delta R^2 = 0.074$).


\section{Markov Transition Matrices}
\label{app:transitions}

Human-derived seven-state transition matrices used for Markov steering (RQ3). Each cell shows $P(\text{column} \mid \text{row})$. The seven states are: \RoleExploring (\AskNur), \RoleSuggesting (\AskAF), \RoleReframing (\AdvNur), \RoleDirecting (\AdvAF), \RoleAffirming (\AckNur), \RoleRelating (\AckAF), and \RoleReflecting (\RBNur). Position-conditioned matrices (early, mid, late) are available in the replication package (\texttt{data/synthetic/transition\_matrices.json}).

\begin{table*}[t]
  \centering
  \caption{Overall transition matrix: MI (AnnoMI, $N = 2{,}083$ segments, 131 conversations).}
  \label{tab:trans-mi}
  \begin{tabular}{@{} l l rrrrrrr @{}}
    \toprule
    & & \multicolumn{7}{c}{To} \\
    \cmidrule(lr){3-9}
    From & Role State & \AskNur & \AskAF & \AdvNur & \AdvAF & \AckNur & \AckAF & \RBNur \\
    &            & Exploring & Suggesting & Reframing & Directing & Affirming & Relating & Reflecting \\
    \midrule
    \AskNur & Exploring  & .463 & .008 & .044 & .057 & .116 & .001 & .311 \\
    \AskAF  & Suggesting & .381 & .000 & .048 & .190 & .143 & .048 & .190 \\
    \AdvNur & Reframing  & .319 & .009 & .115 & .159 & .124 & .009 & .265 \\
    \AdvAF  & Directing  & .189 & .022 & .065 & .427 & .157 & .027 & .114 \\
    \AckNur & Affirming  & .324 & .004 & .061 & .097 & .247 & .004 & .263 \\
    \AckAF  & Relating   & .167 & .000 & .000 & .333 & .500 & .000 & .000 \\
    \RBNur  & Reflecting & .383 & .014 & .066 & .066 & .122 & .003 & .347 \\
    \bottomrule
  \end{tabular}
\end{table*}

\begin{table*}[t]
  \centering
  \caption{Overall transition matrix: Peer (ESConv, $N = 2{,}275$ segments, 250 conversations).}
  \label{tab:trans-peer}
  \begin{tabular}{@{} l l rrrrrrr @{}}
    \toprule
    & & \multicolumn{7}{c}{To} \\
    \cmidrule(lr){3-9}
    From & Role State & \AskNur & \AskAF & \AdvNur & \AdvAF & \AckNur & \AckAF & \RBNur \\
    &            & Exploring & Suggesting & Reframing & Directing & Affirming & Relating & Reflecting \\
    \midrule
    \AskNur & Exploring  & .319 & .015 & .047 & .200 & .319 & .019 & .082 \\
    \AskAF  & Suggesting & .167 & .000 & .042 & .500 & .292 & .000 & .000 \\
    \AdvNur & Reframing  & .081 & .023 & .093 & .349 & .395 & .035 & .023 \\
    \AdvAF  & Directing  & .114 & .009 & .040 & .419 & .355 & .053 & .011 \\
    \AckNur & Affirming  & .135 & .013 & .040 & .251 & .486 & .051 & .024 \\
    \AckAF  & Relating   & .055 & .000 & .014 & .466 & .425 & .041 & .000 \\
    \RBNur  & Reflecting & .358 & .000 & .012 & .210 & .383 & .000 & .037 \\
    \bottomrule
  \end{tabular}
\end{table*}

\begin{table*}[t]
  \centering
  \caption{Overall transition matrix: Counselling (HOPE, $N = 2{,}929$ segments, 158 conversations).}
  \label{tab:trans-counselling}
  \begin{tabular}{@{} l l rrrrrrr @{}}
    \toprule
    & & \multicolumn{7}{c}{To} \\
    \cmidrule(lr){3-9}
    From & Role State & \AskNur & \AskAF & \AdvNur & \AdvAF & \AckNur & \AckAF & \RBNur \\
    &            & Exploring & Suggesting & Reframing & Directing & Affirming & Relating & Reflecting \\
    \midrule
    \AskNur & Exploring  & .553 & .007 & .034 & .071 & .030 & .001 & .303 \\
    \AskAF  & Suggesting & .379 & .034 & .000 & .310 & .069 & .000 & .207 \\
    \AdvNur & Reframing  & .417 & .016 & .071 & .213 & .047 & .000 & .236 \\
    \AdvAF  & Directing  & .315 & .014 & .065 & .283 & .079 & .007 & .237 \\
    \AckNur & Affirming  & .463 & .009 & .111 & .130 & .028 & .000 & .259 \\
    \AckAF  & Relating   & .250 & .000 & .000 & .250 & .250 & .000 & .250 \\
    \RBNur  & Reflecting & .482 & .011 & .046 & .072 & .030 & .000 & .359 \\
    \bottomrule
  \end{tabular}
\end{table*}


\section{Steering Experiment Results}
\label{app:steering}

\begin{table}[h]
  \centering
  \caption{Jensen--Shannon divergence between AI-generated and human reference role distributions.}
  \label{tab:js-div}
  \begin{tabular}{@{} l rrr @{}}
    \toprule
    Condition & MI & Peer & Couns. \\
    \midrule
    Baseline (unsteered) & 0.130 & 0.324 & 0.119 \\
    Markov-steered       & 0.019 & 0.051 & 0.036 \\
    Prompt-only          & 0.096 & 0.358 & 0.211 \\
    \textbf{Markov reduction} & \textbf{85\%} & \textbf{84\%} & \textbf{70\%} \\
    \bottomrule
  \end{tabular}
\end{table}

\begin{table}[h]
  \centering
  \caption{Realisation rates: proportion of turns where the classifier agrees with the Markov chain's sampled role.}
  \label{tab:realisation}
  \begin{tabular}{@{} l rrrrr @{}}
    \toprule
    Domain & Overall & \AskNur & \RBNur & \AckNur & \AdvAF \\
           &         & Exploring & Reflecting & Affirming & Directing \\
    \midrule
    MI          & 67.2\% & 87.8\% & 63.0\% & 51.8\% & 56.4\% \\
    Peer        & 62.8\% & 83.2\% & 50.0\% & 70.5\% & 56.9\% \\
    Counselling & 68.5\% & 88.5\% & 53.7\% & 38.5\% & 38.2\% \\
    \bottomrule
  \end{tabular}
\end{table}

\begin{table}[h]
  \centering
  \caption{Mean consecutive same-role run length. Markov steering reduces run lengths toward human levels; prompt-only steering paradoxically increases rigidity.}
  \label{tab:run-lengths}
  \begin{tabular}{@{} l rrrr @{}}
    \toprule
    Domain & Baseline & Markov & Prompt-only & Human \\
    \midrule
    MI          & 2.11 & 1.59 & 3.17 & ${\sim}$1.5 \\
    Peer        & 3.95 & 1.64 & 3.28 & ${\sim}$1.8 \\
    Counselling & 2.29 & 1.81 & 3.12 & ${\sim}$1.6 \\
    \bottomrule
  \end{tabular}
\end{table}


\section{Steering Decay by Role State}
\label{app:decay}

\begin{table}[h]
  \centering
  \caption{Realisation rate by sampled role and conversational tercile (all domains pooled). Directive roles decay; acknowledgement grows---the model defaults to empathic validation as context fills.}
  \label{tab:decay-by-role}
  \begin{tabular}{@{} l l rrr r @{}}
    \toprule
    Sampled role & Role State & Early & Mid & Late & $\Delta$ \\
    \midrule
    \AskNur & Exploring  & 96.7\% & 87.6\% & 71.5\% & $-25.2$ \\
    \AdvAF  & Directing  & 88.5\% & 46.2\% & 25.0\% & $-63.5$ \\
    \RBNur  & Reflecting & 82.3\% & 46.8\% & 32.9\% & $-49.4$ \\
    \AckNur & Affirming  & 51.7\% & 62.6\% & 73.2\% & $+21.5$ \\
    \AdvNur & Reframing  &  7.7\% & 25.0\% & 28.9\% & $+21.2$ \\
    \bottomrule
  \end{tabular}
  \begin{flushleft}\small Early: turns 0--6; Mid: 7--13; Late: 14--19. $N = 1{,}800$ Markov-condition turns (30 conversations $\times$ 20 turns $\times$ 3 domains). $\Delta$ = late $-$ early in percentage points.\end{flushleft}
\end{table}

A logistic regression of adherence on turn index $\times$ sampled role (combined $N = 1{,}762$, pseudo-$R^2 = 0.228$) confirms these differential trajectories. The main effect of turn is negative ($\beta = -0.171$, $z = -7.31$, $p < .001$), while the \AckNur $\times$ turn interaction is strongly positive ($\beta = +0.231$, $z = 7.52$, $p < .001$), indicating that acknowledgement becomes \textit{more} likely to be produced as conversations progress. Domain-specific models reveal that this attractor is strongest in MI (\AckNur $\times$ turn: $\beta = +0.295$, $z = 5.15$, $p < .001$) and that Peer shows particularly steep decay for \AdvAF ($\beta = -0.285$, $z = -6.17$, $p < .001$).


\section{RQ3 Prompt Templates}
\label{app:prompts}

The three conditions use the following supporter system prompts. The Markov condition prepends a per-turn act instruction (sampled from the transition matrix) to the base prompt; the prompt-only condition prepends the structural guidance block.

\para{MI --- Baseline.} ``You are a trained motivational interviewing (MI) practitioner. Someone has come to you for support with a behaviour they are ambivalent about changing. Use MI principles: express empathy through reflective listening, develop discrepancy, roll with resistance, support self-efficacy. Keep your response to 1--2 sentences (10--25 words). Be brief---say one thing, not three.''

\para{MI --- Prompt-only (structural addition).} ``IMPORTANT STRUCTURAL GUIDANCE: Always reflect what the person has said before offering any direction. Use open questions to explore their experience and develop discrepancy. Roll with resistance---do not argue or confront. Do not give advice or suggestions until the person has expressed readiness to change. When you do offer direction, keep it brief and grounded in what they have shared.''

\para{Peer --- Baseline.} ``You are a peer supporter helping someone through a difficult experience. You are not a therapist---you are a caring person who listens, validates, and shares understanding. Be warm, present, and human. Keep your response to 1--2 sentences (10--25 words). Be brief---say one thing, not three.''

\para{Peer --- Prompt-only (structural addition).} ``IMPORTANT STRUCTURAL GUIDANCE: Acknowledge and validate the person's feelings before anything else. Show that you hear them and that their feelings make sense. If you offer suggestions, do so gently and only after they feel heard. Do not lecture, diagnose, or prescribe. Be a companion, not an authority.''

\para{Counselling --- Baseline.} ``You are a professional counsellor in a therapeutic session. Someone has come to you for support with emotional or psychological difficulties. Use core counselling skills: empathic listening, open questions, reflection, and gentle exploration of feelings. Follow the client's lead---do not impose an agenda. Help them hear what they are saying, not what you think they should hear. Keep your response to 1--2 sentences (10--25 words). Be brief---say one thing, not three.''

\para{Counselling --- Prompt-only (structural addition).} ``IMPORTANT STRUCTURAL GUIDANCE: Begin by exploring and reflecting---ask open questions and mirror back what the client shares. Use reflective listening extensively: paraphrase, summarise, name emotions the client has implied but not stated. Acknowledge is rare in therapeutic counselling---avoid generic validation; instead, show understanding through specific reflection. Only offer interpretive or directive input after substantial exploration.''


\end{document}